\documentclass[11pt,a4paper]{article}
\usepackage{cite}
\usepackage{amsmath, amsthm, amssymb,slashed,upgreek,textgreek,url}
\usepackage{ifpdf}
\ifpdf
  \usepackage[pdftex]{graphicx}
  \usepackage{epstopdf}
\else 
  \usepackage[dvips]{graphicx}
\fi
\parskip=\baselineskip
\def\Bbb{\mathbb}

\def\16{{\bf 16}}
\def\1{{\bf 1}}
\def\2{{\bf 2}}
\def\3{{\bf 3}}
\def\4{{\bf 4}}
\def\8{{\bf 8}}
\def\sF{{\sf F}}
\def\z{z}
\def\bg{{\overline\gamma}}
\def\mc{{\mathrm{m}}_c}
\def\T{{\mathcal T}}
\def\h{\widehat}
\def\bG{\overline\Gamma}
\def\F{{\mathcal F}}
\def\q{{\sf q}}

\def\K{{\mathcal K}}
\def\mod{{\mathrm{mod}}}
\def\S{{\mathcal S}}
\def\o{{\mathfrak o}}

 \def\Spin{{\mathrm{Spin}}}
 \def\SU{{\mathrm{SU}}}
 \def\SO{{\mathrm{SO}}}
 
 \def\spinc{\mathrm{spin}_c}
 \def\veps{\varepsilon}
 \def\mo{{\mathfrak o}}
 \def\vepsk{{\mathfrak o}^{\oplus k}}
\def\bar{\overline}

\def\R{{\Bbb{R}}}\def\Z{{\Bbb{Z}}}
\def\ZZ{{\mathcal Z}}

\def\hat{\widehat}
\font\teneurm=eurm10 \font\seveneurm=eurm7 \font\fiveeurm=eurm5
\newfam\eurmfam
\textfont\eurmfam=\teneurm \scriptfont\eurmfam=\seveneurm
\scriptscriptfont\eurmfam=\fiveeurm

\font\teneusm=eusm10 \font\seveneusm=eusm7 \font\fiveeusm=eusm5
\newfam\eusmfam
\textfont\eusmfam=\teneusm \scriptfont\eusmfam=\seveneusm
\scriptscriptfont\eusmfam=\fiveeusm

\font\tencmmib=cmmib10 \skewchar\tencmmib='177
\font\sevencmmib=cmmib7 \skewchar\sevencmmib='177
\font\fivecmmib=cmmib5 \skewchar\fivecmmib='177
\newfam\cmmibfam
\textfont\cmmibfam=\tencmmib \scriptfont\cmmibfam=\sevencmmib
\scriptscriptfont\cmmibfam=\fivecmmib

\numberwithin{equation}{section}

\def\d{\mathrm d}

\def\Z{{\Bbb Z}}
\def\th{\mathrm{th}}

\def\bar{\overline}
\def\RP{{\Bbb{RP}}}

\def\I{{\mathcal I}}
\def\d{{\mathrm d}}

\def\CP{\Bbb{CP}}
\def\R{{\Bbb R}}

\def\Z{{\Bbb Z}}
\def\slT{\text{\sffamily\slshape T\/}}
\def\sK{{\sf K}}
\def\Yuk{{\mathrm{Yuk}}}
\def\eff{{\mathrm{eff}}}

\def\L{{\mathcal L}}

\def\t{{\mathfrak t}}

\def\d{{\mathrm d}}
\def\R{{\Bbb R}}
\def\Z{{\Bbb Z}}

\def\H{{\mathcal H}}
\def\ti{\widetilde}

\def\L{{\mathcal L}}
\def\D{{\mathcal D}}

\def\Pf{{\mathrm{Pf}}}
\def\be{\begin{equation}}
\def\ee{\end{equation}}

\def\sRT{\sf{RT}}

\begin{document}
\begin{titlepage}
\begin{flushright}

\end{flushright}
\vskip 1.5in
\begin{center}
{\bf\Large{The ``Parity'' Anomaly}}
{\bf\Large{ on an Unorientable Manifold}}
\vskip
0.5cm {Edward Witten} \vskip 0.05in {\small{ \textit{School of
Natural Sciences, Institute for Advanced Study}\vskip -.4cm
{\textit{Einstein Drive, Princeton, NJ 08540 USA}}}
}
\end{center}
\vskip 0.5in
\baselineskip 16pt
\begin{abstract} The ``parity'' anomaly -- more accurately described as an anomaly in time-reversal 
or reflection symmetry -- arises in certain theories
of fermions coupled to gauge fields and/or gravity in a spacetime of odd dimension.  This anomaly 
has traditionally been studied on orientable manifolds
only, but recent developments involving topological superconductors have made it clear that one can get more 
information by asking what
happens on an unorientable manifold. In this paper, we give a full description of the ``parity'' anomaly for fermions 
coupled to gauge fields and gravity in $2+1$ dimensions on a possibly unorientable spacetime.   We consider an application to topological superconductors
 and another application to M-theory.  The application to topological superconductors involves using  knowledge
 of the ``parity'' anomaly as an ingredient in constructing gapped boundary states of these systems and in particular
 in gapping the boundary of a $\nu=16$ system in a topologically trivial fashion.  The application to M-theory involves
 showing the consistency of the path integral of an M-theory membrane on a possibly unorientable worldvolume.
 In the past, this has been done only in the orientable case.

  \end{abstract}
\date{March, 2016}
\end{titlepage}
\def\Hom{\mathrm{Hom}}
\def\U{{\mathrm U}}
\def\SU{{\mathrm{SU}}}
\def\sT{{\sf T}}
\def\sR{{\sf R}}
\def\sC{{{\sf C}}}
\def\OO{{\mathrm O}}
\def\sCT{{\sf {CT}}}
\def\sCR{{\sf {CR}}}
\def\g{\gamma}
\def\G{\Gamma}
\def\i{{\mathrm i}}
\def\D{{\mathcal D}}
\def\sCRT{{\sf{CRT}}}
\def\sCPT{\sf{CPT}}
\def\pin{{\mathrm{pin}}}
\tableofcontents
\section{Introduction}

A $\U(1)$ gauge theory in three spacetime dimensions, coupled to a single massless Dirac fermion $\chi$ of charge 1, is invariant at the classical
level under time-reversal and reflection symmetry, which we will call $\sT$ and $\sR$.  However, quantum mechanically there is an anomaly: to quantize this theory in a gauge-invariant
fashion, one must give up $\sT$ and $\sR$ symmetry \cite{Redlich, Semenoff, ADM}.  This anomaly is commonly called a 
``parity'' anomaly, but this terminology is somewhat
misleading as parity (a spatial inversion, acting as $-1$ on all spatial coordinates) is in the connected 
component of the rotation group in $2+1$ dimensions.
The anomaly is better understood as an anomaly in $\sT$ and $\sR$.

In this paper, we consider a refinement of the usual ``parity'' anomaly. We will make this analysis for arbitrary gauge groups,
but for gauge group $\U(1)$, what we will learn can be stated as follows.
   The usual statement of the anomaly is that  a $\U(1)$ gauge theory in $2+1$
dimensions coupled to a single Dirac fermion $\chi$ of charge 1 cannot be quantized in a gauge-invariant way, consistent with $\sR$ and 
$\sT$ symmetry.  If one has two such  Dirac fermions, both of charge 1, then the
usual ``parity'' anomaly is absent and, on an orientable manifold, the theory can be quantized in a way that preserves $\U(1)$ gauge-invariance
as well as  $\sR$ and $\sT$.  If $\sR$ and $\sT$ were ordinary global symmetries (not acting on spacetime), one would now ask: Can these symmetries be gauged?  In general
for ordinary global symmetries, the answer to such a question can be ``No'':  even if the symmetries are valid as global symmetries, there
may be an anomaly (sometimes called an 't Hooft anomaly) that would obstruct gauging them.   For $\sR$ and $\sT$, the closest analog
of gauging the symmetry is to use these symmetries to formulate a theory on an unorientable manifold.\footnote{If a theory with $\sR$ or $\sT$ symmetry can be formulated on
 an unorientable manifold, we say that $\sR$ or $\sT$ is gaugeable, while actually gauging the symmetry in the context of quantum gravity would mean summing over unorientable manifolds.}  Therefore, to get a refined
version of the usual ``parity'' anomaly, we can consider a $\U(1)$ gauge theory with a general number $y$ of Dirac fermions of charge 1,
and ask for what values of $y$ this theory can be placed on an unorientable manifold without violating gauge-invariance.
We shall answer this question and find that $y$ must be a multiple of 4.  Thus while the usual ``parity'' anomaly is a mod 2 effect,
its refinement in which one ``gauges'' $\sR$ and $\sT$ symmetry is really a mod 4 effect.

The existence of this refinement should not come as a surprise, since the purely gravitational analog is already known.\footnote{Also,
it is already known from another point of view (see section IIIC of \cite{WS} and Appendix  C of \cite{MCFV}) that a purely $2+1$-dimensional $\sT$-invariant
$\U(1)$ gauge theory with $4n+2$
Dirac fermions of charge 1 is anomalous 
from the standpoint of condensed matter physics.  The basic monopole operator of this theory, defined by a monopole singularity of charge 1,
is a bosonic operator that transforms under $\sT$ as a Kramers doublet.  This is not possible in a purely $2+1$-dimensional system 
that is built microscopically from electrons and nuclei. We discuss this issue in section \ref{monop}.}  Consider a theory
of $\nu$ Majorana fermions, coupled to gravity only.  As long as one is on an orientable manifold, the fermion path integral of this system is real in Euclidean
signature (because the Dirac operator is hermitian).  One should worry about a possible problem in defining the {\it sign} of the fermion
path integral.  However, as long as $\nu$ is even -- to ensure that the path integral is positive -- the theory of $\nu$ Majorana fermions
is completely well-defined and $\sR$- and $\sT$-invariant on an orientable manifold.\footnote{\label{howth}  In this introduction,  we take all $\nu$ fermions $\psi_1,\dots,\psi_\nu$ to transform in the same way under $\sT$ or $\sR$.
Specifically, we define $\sT$
by $\sT\psi_i(t,\vec x)=\g_0 \psi_i(-t,\vec x)$, $i=1,\dots,\nu$, where here $\g_\mu$, $\mu=0,\dots,2$ are the Dirac gamma matrices, and similarly
for $\sR$.  A fermion transforming 
with the opposite sign -- as $\sT\psi(t,\vec x)=-\g_0 \psi(-t,\vec x)$ -- would make a contribution to the anomaly with the opposite sign.
 In general, $\nu$ should be defined as the number of fermions
 that transform with a $+$ sign minus the number that transform with a $-$ sign.
 For more on this, see \cite{Witten}, as well as  section \ref{analysis} below. }
 Can the $\sR$ and $\sT$ symmetries be gauged,
or more precisely is the theory well-defined when formulated on an unorientable manifold?   The answer to this question is that when the
theory of $\nu$ Majorana fermions is formulated on an unorientable manifold, one runs into an anomaly that involves an eighth root of 1
in the simplest calculations \cite{R}, 
and a sixteenth root of 1  in a more systematic analysis (see section 4 of \cite{Witten}).   

The mod 16 anomaly for $2+1$-dimensional
Majorana fermions coupled to gravity only is intimately related to the theory of a topological superconductor in $3+1$ dimensions.
A topological superconductor in $3+1$ dimensions is characterized by an invariant usually called $\nu$;  in Euclidean signature, it
has a bulk partition function $\exp(-\nu\pi\i\eta/2)$, where $\eta$ is the eta-invariant of the four-dimensional Dirac operator.  One can think of this
factor, which was suggested in \cite{Ketal}, as coming from integrating out $\nu$ bands\footnote{Or more generally any number of bands with a net invariant of $\nu$.}
 of  gapped bulk fermions  \cite{Witten,Metlitski}.   Here $\exp(-\pi\i\eta/2)$
is a topological invariant if the worldvolume $X$ of the topological superconductor has no boundary; 
it is the partition function of a topological field theory defined on closed four-dimensional
spin (or more precisely $\pin^+$) manifolds.  This invariant is in general an arbitrary 
16$^{\th}$ root of 1, so $\exp(-\nu\pi\i\eta/2)$ is trivial (for all $X$) if and only
if $\nu$ is divisible by 16. But what happens if $X$ has a boundary? Then, if $\nu$ is not a multiple of 16, $\exp(-\nu\pi\i\eta/2)$ cannot be defined 
 as a topological invariant.
However, if we include $\nu$ massless Majorana fermions  on the boundary  of $X$ (and transforming under $\sT$ as
in footnote \ref{howth}),  the
anomaly of the fermions compensates for the ill-definedness of the bulk partition function and the combined system is well-defined and anomaly-free.  In this paper, we will analyze the generalization of this construction to include gauge fields -- either gauge fields that propagate
in $3+1$ dimensions, or gauge fields that propagate only on the boundary of a four-manifold.

In band theory, and as we have defined it above, 
$\nu$ is an integer, but when fermion interactions in the bulk theory are taken into account, it is known from several points of view \cite{FCV,WS,MCFV,KitTwo,Tach}
that $\nu$ is only an invariant mod 16.  One way to understand this is the fact that the partition function of the bulk theory and the anomaly of
the boundary theory depend only on the value of $\nu$ mod 16.  

This paper is organized as follows.
We analyze the general
  ``parity'' anomaly for gauge fields on an unorientable manifold\footnote{Mathematical background to this rather technical analysis can be found in \cite{Witten}.
The applications in the rest of the paper can be understood without a detailed reading of section \ref{analysis}, provided the reader is willing to take a few statements on faith.} in section \ref{analysis}.   
In the remainder of the paper, we describe two applications of the result, one in condensed matter physics and one in string/M-theory.

In section \ref{gb}, we use our results to construct gapped, symmetry-preserving boundary states of a topological superconductor.
Such states and analogous ones for topological insulators and for certain bosonic states of matter have been analyzed in the literature from various points of view\cite{WS,MCFV,MKF,TPfaffian,TPfaffiantwo,Wangetal,Wangetals,MV,Al,VS,MetlitskiTSC}.
We use here the same methods   as in \cite{SW} and our basic examples
were already analyzed there.  However, having a more complete knowledge of which models are anomaly-free enables us to be more precise
on some points, and in particular, it will help us to construct in a straightforward way a gapped and topologically trivial boundary state at $\nu=16$.
(If one incorrectly believed that the ``parity'' anomaly is a mod 2 effect even on an unorientable manifold, one would incorrectly
infer the existence of a gapped and topologically trivial boundary state at $\nu=8$.)
We primarily consider models that are gapped at the perturbative level, but in section \ref{confinement},
we consider models  that are gapped with the help of nonperturbative effects.
Here we give two related constructions,  both at $\nu=16$, one of which uses the mass gap of  $\SU(2)$ gauge theory and one of 
which uses the Polyakov model \cite{Polyakov} of compact
QED in $2+1$ dimensions.   (An interpolation between different bulk states at $\nu=16$ using nonperturbative properties of four-dimensional
gauge theory has been described in \cite{Tach}.)

In section \ref{mtheory}, we apply some of the same ideas to a problem in string/M-theory.  It has been known \cite{Wittenold}
that the ``parity'' anomaly in $2+1$ dimensions
has implications for M2-branes (membranes with $2+1$-dimensional worldvolume that arise in M-theory).  In fact, the fermions that propagate on the M2-brane
worldvolume do have an anomaly, which has to be canceled by a version of anomaly inflow \cite{CH} involving  a
shifted quantization law of the three-form field of M-theory.  In the past,
this analysis has been made only for the case that the M2-brane worldvolume is orientable.  Here we analyze the unorientable case,
showing that, with a slightly refined statement of the shifted quantization law, anomalies still cancel.  (Previously, the path integral on an unorientable string
worldsheet has been analyzed in \cite{DFM2,DFM}.)  A short summary of the paper
can be found in section \ref{conclusions}.

We conclude this introduction with a general comment.  
Apart from technical details, the analysis in this paper
 differs  in the following way from more familiar discussions of the ``parity'' anomaly.
On an orientable manifold, the usual ``parity'' anomaly is a conflict between $\sT$ or $\sR$
symmetry and gauge invariance.  However, when we use  $\sR$ symmetry to formulate a theory on an unorientable manifold 
(we work in Euclidean signature so $\sR$ is more relevant here than $\sT$), that
symmetry is built in and it is too late
to give it up.  So the anomalies we compute on unorientable manifolds represent breakdowns of gauge symmetry or diffeomorphism
symmetry rather than conflicts between those local symmetries and $\sT$ or $\sR$ symmetry.

\section{Analysis Of The Anomaly}\label{analysis}

\subsection{Preliminaries}\label{preliminaries}

In this paper, we will consider fermions on a $2+1$-dimensional manifold $W$ interacting with gauge fields of some gauge group $G$ and with gravity.  
In the main condensed matter application, $W$ will be the boundary of  a four-manifold $X$ that will be the worldvolume  of a topological superconductor and
$G$ will be a group of emergent gauge fields that propagate only along $W$ (and not  $X$).  In the string/M-theory example, 
$W$ will be the worldvolume of an M2-brane,
and $G$ will be the structure group of the normal bundle to $W$ in the full higher-dimensional spacetime.

In either case, we consider gauge fields $a$ that transform under time-reversal $\sT$ in the natural-looking way 
\begin{align}\label{transf}\sT (a_0(x^0,\vec x))=& -a_0(-x^0,\vec x)\cr
                                       \sT (a_i(x^0,\vec x)) = & \,a_i(-x^0,\vec x),~~i=1,2. \end{align}
In terms of differential forms, if $\tau$ is the time-reversal transformation of spacetime  $x^0,\vec x\to -x^0,\vec x$, eqn. (\ref{transf}) would
be written  $\sT(a)=\tau^*(a)$.  These conditions imply that the gauge charge (which couples to $a_0$) is odd
under $\sT$, and the ``magnetic'' flux (the integral of $F_{12}=\partial_1 a_2-\partial_2a_1$) is even.  This
behavior under $\sT$ will be assumed for all gauge fields considered in the present paper.

The choice in eqn. (\ref{transf}) really requires some explanation.  In nature, there is also an operation $\sC$ of charge conjugation, so there are two possible time-reversal operations,
namely $\sT$ and $\sCT$.  The usual convention in particle physics and condensed matter physics
is that the gauge field $A$ of electromagnetism transforms under $\sT$ with an opposite sign to eqn. (\ref{transf})
-- thus $\sT(A)=-\tau^*(A)$, as opposed to $\sCT(A)=+\tau^*(A)$.   This definition ensures that electric charge is $\sT$-even.
By contrast, the  choice (\ref{transf}) means that the conserved charge $\q$ that couples to $a_0$ is $\sT$-odd.  

A rationale for the choice (\ref{transf}) is that in the present paper, we will consider problems in which there is no natural operation
corresponding to $\sC$.    Since $\sC$ reverses the sign of the electric charge, and electric charge is not conserved in a 
superconductor,\footnote{To be more precise, in a superconductor electric charge is conserved mod 2, meaning that the sign of the charge
is irrelevant and thus there is no natural $\sC$ operation.  To say it differently, a superconductor has fermionic quasiparticles
but there is not a meaningful sense in which their electric charge is positive or negative.  In terms of the electromagnetic
gauge field $A$, one would say that in a superconductor, as $U(1)$ is broken to $\Z_2$, $A$ becomes a $\Z_2$ gauge
field, so that $2A$ is gauge-equivalent to 0 at long distances and the operation $\sC:A\to -A$ is irrelevant.  From a macroscopic point of view, this is true even in the presence of flux lines.  A fermionic excitation propagates around a flux
line with a change of sign regardless of whether the flux line has flux $\pi$ or $-\pi$.}
 there is no natural
$\sC$ operation in the theory of a topological superconductor. Likewise, in the string/M-theory problem, there is no  operation corresponding to $\sC$.

Since there will be no analog of $\sC$ in the problems that we will consider, we will simply refer to the transformation (\ref{transf}) as $\sT$.
Actually, to analyze anomalies, which will be our goal in the present section, it is convenient to work in Euclidean signature.  Here we consider
a spatial reflection rather than time-reversal.  We take $\sR$ to act by reflection of one coordinate
\begin{align}\label{transtwo}\sR (a_1(x_1,x_2,x_3))=& -a_1(-x_1,x_2,x_3)\cr
                                       \sR (a_i(x_1,x_2,x_3)) = &\, a_i(-x_1,x_2,x_3),~~i>1 .\end{align}
 Equivalently, $\sR(a)=\rho^*(a)$, where $\rho$ is the spatial reflection $x_1,x_2,x_3\to -x_1,x_2,x_3$.                                      

Thus, in going to an unorientable manifold, we will consider gauge fields $a$ that transform under an orientation-reversing diffeomorphism $\rho$
as $a\to \rho^*(a)$.   As we will see, this leads to a richer structure than the familiar ``parity'' anomaly that is seen on an orientable
manifold.  By contrast, if one uses $\sCR $ to go to an unorientable manifold (for example, by considering a $U(1)$ gauge field $a$
that transforms under an orientation-reversing diffeomorphism as $a\to -\rho^*(a)$), then one finds only the standard ``parity'' anomaly.
This has been explained in section 4.7 of \cite{Witten}.   That is why the considerations of the
present paper are more directly relevant to topological superconductors (and  M-theory)
 rather than to topological insulators.
 
We reiterate that in this paper, we consider theories that have no natural $\sC$ operation 
at the microscopic level so the discrete symmetries are $\sT$ and $\sR$.   Should we contemplate the possibility of $\sC$ as an emergent
symmetry?  In general, {\it exact} emergent symmetries are
gauge symmetries; an emergent $\sC$ symmetry, if it is an exact symmetry, would be  part
of the emergent gauge group and could be incorporated as such in the analysis that follows.  It is certainly possible to have {\it approximate}
emergent global symmetries, but an approximate $\sC$ symmetry would not be relevant in the present paper; for example,
we cannot use $\sCR $ in putting a theory on an unorientable manifold unless $\sCR$ is an exact symmetry.

A final comment is that  what we call $\sT$ and $\sR$ in this paper correspond more closely to  what are usually called
$\sCT$ and $\sR$, respectively, when electric charge conservation is relevant and  in theories with a 
microscopic $\sC$ symmetry.\footnote{\label{notational} For example, what we call $\sT$ was called $\sCT$ in sections 2-5 of \cite{SW}, which are devoted
to a topological insulator (in which electric charge is conserved).  It was called $\sT_{\mathrm{sc}}$ in section 6 of that paper, which is 
devoted to a topological superconductor.}   In a theory that does have a microscopic $\sC$ symmetry, analytic continuation from Lorentz
signature to Euclidean signature relates $\sCT$ to $\sR$ and $\sT$ to $\sCR$.   

\subsection{The Fermion Path Integral}\label{condensed}

In general, there is a choice of sign in the transformation of a $2+1$-dimensional fermion field $\psi$ under $\sT$:
\begin{align}\label{zilb}\sT(\psi(t,\vec x))=\pm \g_0 \psi(-t,\vec x). \end{align}
We note that for either choice of sign, one has $\sT^2=(-1)^\sF$ (here $(-1)^\sF$ is the operator that counts the number of fermions mod 2).
However, the two choices are physically inequivalent; no linear transformation of $\psi$ would reverse this sign.

There is a similar sign choice in the transformation under a spatial reflection.  In Euclidean signature
\be\label{ilb}\sR(\psi(x_1,x_2,x_3))=\pm \g_1\psi(-x_1,x_2,x_3). \ee
In a relativistic theory, the signs in eqns. (\ref{zilb}) and (\ref{ilb}) can be equated, as the $\sCRT$ theorem\footnote{Here $\sC$ can be dropped if gauge fields
transform under $\sT$ and $\sR$ as stated in eqns. (\ref{transf}) and (\ref{transtwo}).  What we are calling $\sT$ is called $\sCT$ in the usual
statement of the theorem, so what is usually called $\sCRT$ reduces to $\sRT$ with our choices. 
As explained in section \ref{preliminaries}, we are using conventions that are more natural in a context
(such as the theory of a topological superconductor) in which there is no natural $\sC$ operation.} ensures that a theory has a $\sT$ symmetry
with a given choice of sign if and only if it has an $\sR$ symmetry with the same choice of sign.

We consider a system of $2+1$-dimensional massless Majorana 
fermions $\Psi=(\psi_1,\psi_2,\dots,\psi_n)$ transforming in an $n$-dimensional real\footnote{The representation is real because the global 
symmetry group of $n$ massless Majorana fermions in $2+1$ dimensions is ${\mathrm O}(n)$.  $G$ is necessarily a 
subgroup of ${\mathrm O}(n)$ and the representation $R$
is the vector representation of ${\mathrm O}(n)$, restricted to $G$.  This is a real representation because the 
vector representation of ${\mathrm O}(n)$ is real.} representation
$R$ of a gauge group $G$, and all transforming under $\sT$ (or $\sR$) with a + sign.    
We want to know whether this theory makes sense as a purely $2+1$-dimensional theory, and to measure
the anomaly if it does not.  We will schematically call that anomaly $\alpha_R$.   One can think of this anomaly as the change in phase
of the path integral under a diffeomorphism and/or gauge transformation, though we also come later to a more powerful way to think about it involving
topological field theory.

Alternatively, we could assume that the fermions $\Psi$ transform under $\sT$ and $\sR$ with a minus sign.  This complex conjugates
the path integral, so it changes the sign of the anomaly, replacing $\alpha_R$ with $-\alpha_R$.

More generally, we could have one set of fermions $\Psi=(\psi_1,\psi_2,\dots,\psi_n)$ transforming in the representation 
$R$ of $G$, and transforming under $\sT$ with a + sign, and a second set $\Psi'=(\psi'_1,\psi'_2,\dots,\psi'_{n'})$ 
transforming in some possibly different representation
$R'$ of $G$, and transforming under $\sT$ with a $-$ sign.   The net anomaly is then $\alpha_R-\alpha_{R'}$.  
Clearly, then, to understand the anomaly for an arbitrary system of fermions, it suffices to compute $\alpha_R$, that is, it suffices
to consider a system of fermions $\Psi$ that all transform under $\sT$ with a $+$ sign.

Formally, the path integral $Z_\Psi$ of a system of fermions, with Dirac operator $\D=\i\slashed{D}$, is the Pfaffian $\Pf(\D)$.
Our approach to computing the anomaly will be based on the Dai-Freed theorem \cite{DF}.  This theorem\footnote{Some aspects
of the Dai-Freed theorem were recently described for physicists in \cite{Yonek}.} gives a recipe (in Euclidean signature)
to define $Z_\Psi$ in a gauge-invariant fashion, compatible with general physical principles such as unitarity,
once  some auxiliary choices are made.  One then has to check whether the result does depend on the choices.
If it does not, then one has succeeded in 
finding a satisfactory definition of $Z_\Psi$ and the $2+1$-dimensional theory is anomaly-free.  If $Z_\Psi$ as defined by the theorem does depend
on the auxiliary choices, then the theory is anomalous, and its anomaly can be measured by how the theory depends on the choices.

Let $W$ be a compact three-manifold endowed with a $\pin^+$ structure\footnote{A $\pin^+$ structure
is the generalization of a spin structure on a possibly unorientable manifold, for fermions with $\sT^2=(-1)^\sF$.  
See Appendix A of \cite{Witten} for an introduction.}    and a background gauge field.
Suppose that we wish to calculate the path integral $Z_\Psi$ of the fermions $\Psi$ on a possibly unorientable three-manifold $W$
of Euclidean signature.
The Dai-Freed theorem gives us a definition of $Z_\Psi$ once we are given a four-dimensional  manifold
$X$ with boundary $W$ such that the $\pin^+$ structure and the metric and gauge field on $W$ all extend over $X$.
The definition is
\be\label{torof}Z_\Psi =|\Pf(\D)|\exp(-\pi\i\eta_R/2). \ee
Here $\eta_R$ is the Atiyah-Patodi-Singer \cite{APS} eta-invariant of the Dirac operator acting on four-dimensional Majorana fermions
in the representation $R$ of $G$, and $|\Pf(\D)|$ is simply the absolute value of $\Pf(\D)$ (this absolute value is always 
anomaly-free for an arbitrary fermion system).  

Mathematically, the justification for the formula is as follows. First, the right hand side is well-defined and gauge-invariant.  Second, the variation of the right hand side is as expected: that
is, its variation in a change in the background metric or gauge field on $X$ depends only on the fields along $W$
and is  given by the (regularized) one-point function of the stress tensor or current in the fermion theory on $W$, as expected. An important consequence of the second
point is that the
right hand side  vanishes precisely when $Z_\Psi$ is expected to vanish, namely for metrics and gauge fields such
that $\D$ has a zero-mode; and moreover, it
  varies smoothly near such a point in field space (this is required in order for the formula to be physically sensible, and is nontrivial as neither $|\Pf(\D)|$
nor $\exp(-\pi\i\eta_R/2)$ varies smoothly near such a point).  Finally, in view of the gluing theorem for the eta-invariant \cite{DF}, which
will be described below,
the right hand side  has a behavior under cutting and pasting that is compatible with physical
principles such as unitarity.   

Physically, one might interpret the formula (\ref{torof}) as follows, by analogy with standard constructions in 
topological band theory \cite{Kane,SC,Be}.  Consider on $X$ a gapped system with $G$ symmetry consisting of
massive Majorana fermions $\Upsilon$ transforming in the representation $R$ of $G$.  For one sign\footnote{In topological band theory,
the description in terms of a massive Majorana fermion is valid only near a point in the Brillouin zone at which the band gap -- controlled in relativistic terminology by a fermion mass parameter -- is small.
Which sign of the mass parameter corresponds to a trivial theory and which corresponds to a nontrivial theory 
depends on the topology of the rest of the band. In a relativistic theory, defined with Pauli-Villars regularization, 
the massive fermions $\Upsilon$ are accompanied by regulator fields $\widetilde\Upsilon$ of opposite statistics.  
The trivial case is that $\Upsilon$ and $\widetilde\Upsilon$ have mass parameters of the same sign, and the 
nontrivial case is that the mass parameters have opposite signs.} of the $\Upsilon$ mass parameter, this theory is topologically
trivial and can be given a boundary condition such that the theory remains gapped along the boundary.  For the opposite sign of the $\Upsilon$ mass
parameter, the bulk theory is topologically nontrivial, and gapless $2+1$ dimensional Majorana fermions -- with 
the properties of the fields that we have called $\Psi$
-- appear on the boundary.    The formula  (\ref{torof}) for $Z_\Psi$ should be understood as the partition function for the combined system
comprising the gapped fermions $\Upsilon$ in bulk and the gapless fermions $\Psi$ on the boundary.

On a compact four-manifold $X$ without boundary, the $\Upsilon$ path integral (after removing nonuniversal local terms)
would be $Z_\Upsilon=\exp(-\pi\i\eta_R/2)$.  This fact goes back in essence to \cite{ADM} 
and has been exploited in several recent papers \cite{Witten,Metlitski}.  
The right hand side is a topological invariant (invariant, that is, under continuous variation of the
metric and gauge field on $X$).  Indeed, the APS index theorem \cite{APS} implies\footnote{See for example \cite{Witten}, especially
section 2 and Appendix A, for an explanation of assertions made here.  In brief, in even dimensions, $\eta_R$ is a topological invariant
except that it jumps by $\pm 2$ when an eigenvalue of the Dirac operator passes through 0.  In four dimensions, for a real representation $R$ (which
we have here since the $2+1$-dimensional fermions that we started with are always in a real representation of $G$), the eigenvalues of the Dirac
operator have even multiplicity because of a version of Kramers doubling. So the jumps are by $\pm 4$ and  $\eta_R$  is a topological
invariant mod 4.}
 that $\eta_R$ is a topological invariant mod 4, so that $\exp(-\pi\i\eta_R/2)$ is a topological invariant.   In fact $\exp(-\pi\i\eta_R/2)$
is not just a topological invariant, but a cobordism invariant (it equals 1 if $X$ is the boundary of a five-dimensional $\pin^+$ manifold over
which the $G$-bundle over $X$ extends).  This again follows from the APS index theorem.   Cobordism invariance implies\footnote{This assertion is
closely related to the gluing theorem \cite{DF} for the eta-invariant, which we state shortly.} \cite{FH}
that $\exp(-\pi\i\eta_R/2)$ is the partition function of a topological field theory (defined on a four-dimensional $\pin^+$ manifold endowed
with a $G$-bundle).  Cobordism invariance of the eta-invariant in four dimensions motivated the conjecture in \cite{Ketal} that this function would be 
relevant to fermionic SPT phases.

However, if $X$ is a manifold with nonempty boundary $W$, then it is not possible to define $\eta_R$ to make $\exp(-\pi\i\eta_R/2)$ a topological
invariant.  This means that there is a problem in defining on a manifold with boundary the topological field theory that on a manifold without
boundary has partition function $\exp(-\pi\i\eta_R/2)$.  This problem does not have a unique solution; the topological field theory in question
has different possible boundary states.  However, it always has one simple gapless boundary state, in which Majorana fermions in the representation
$R$ propagate on the boundary.   $Z_\Psi=|\Pf(\D)|\exp(-\pi\i\eta_R/2)$ is the partition function of the theory on $X$ with this boundary state along
$W$.  Here, $\eta_R$ is defined using APS boundary conditions.  On a four-manifold without boundary, $\exp(-\pi\i\eta_R/2)$ is a topological invariant;
on a four-manifold with boundary,
 it is not a topological invariant, but varies in such a way that it makes sense in conjunction
with the boundary fermions in the combination $|\Pf(\D)|\exp(-\pi\i\eta_R/2)$.

Now we can discuss whether the theory of the $2+1$-dimensional massless fermions $\Psi$ on $W$ makes sense as a purely $2+1$-dimensional
theory.  In the present framework,  this amounts to asking whether the formula $Z_\Psi=|\Pf(\D)| \exp(-\pi\i\eta_R/2)$ depends on
the choices that were made in defining $\eta_R$.
 There is a standard way to answer this sort of question.  Let $X$ and $X'$ be two different four-manifolds
with boundary $W$, and  with  choices of extension over $X$ and over $X'$ of the $\pin^+$ structure of $W$ and the $G$-bundle over $W$.
Let $\eta_R^X$ and $\eta_R^{X'}$ be the eta-invariants computed on $X$ or on $X'$.  To decide whether the formula 
for $Z_\Psi$ depends on $X$, we  need to know whether $\exp(-\pi\i\eta_R^X/2)   $ is always equal to $\exp(-\pi\i\eta_R^{X'}/2)$.

\begin{figure}
 \begin{center}
   \includegraphics[width=3in]{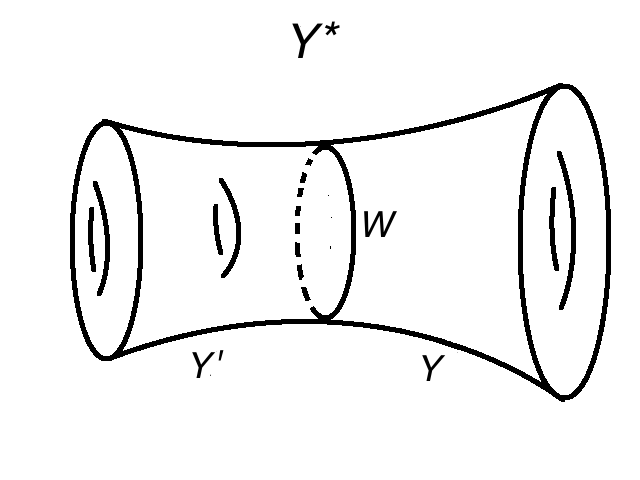}
 \end{center}
\caption{\small  Two manifolds $Y$ and $Y'$ glued along a component of their common boundary to make a manifold $Y^*$ that itself
may have a boundary.}
 \label{second}
 \end{figure}

\begin{figure}
 \begin{center}
   \includegraphics[width=4.5in]{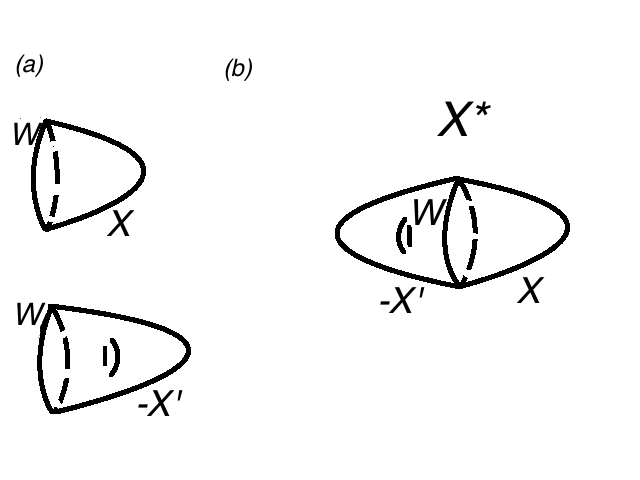}
 \end{center}
\caption{\small  Two manifolds $X$ and $X'$ with common boundary $W$ are glued together along their common boundary
to build a compact manifold $X^*$ without boundary.}
 \label{first}
 \end{figure}

The gluing
theorem for the eta-invariant \cite{DF} gives a convenient way to answer this question.  
To motivate the statement of this theorem, let us first recall the following basic fact about local effective actions in quantum field theory.
Let $I^Y_\eff=\int_Y\d^4x \sqrt g {\mathcal L}$
be any local effective action on a manifold $Y$.  Let $Y$ and $Y'$ be two manifolds with boundary such that $W$ is a boundary component of each.
Suppose further that $Y$ and $Y'$ can be glued together along $W$ to make a manifold\footnote{$W$ may be only one component of the boundary
of $Y$ and $Y'$, so that $Y^*$ may be itself a manifold with boundary, as in the figure. However, the boundary of $Y^*$ will be empty
 in the application we make momentarily.}  $Y^*$, as in fig. \ref{second}.  We assume that in this gluing, all structures on $Y$ and $Y'$
that are used in defining the effective action are compatible. Thus, if $Y$ and $Y'$ are orientable (and the definition of the action depends on the orientation)
the orientations must agree in the gluing; if they are spin manifolds, the spin structures must agree, and if they are $\pin^+$ manifolds the $\pin^+$ structures must agree.
Under these conditions, for any local effective action, we have 
the gluing or factorization property
\be\label{luth}\exp(\i I^Y_\eff)\exp(\i I^{Y'}_\eff)=\exp(\i I_\eff^{Y^*}) \ee
which is intimately connected with locality and unitarity in physics.
The gluing theorem for the eta-invariant says that even though $\eta_R$  is not the integral of a local effective action, $\exp(-\pi\i\eta_R/2)$ behaves as if it were:
\be\label{gluth}{\exp\biggl(-\pi\i\eta_R^Y/2\biggr)}{\exp\biggl(-\pi\i \eta_R^{Y'}/2\biggr)} =\exp\biggl(-\i \pi \eta_R^{Y^*}/2\biggr). \ee 
It is because of this formula that it is physically sensible to have a topological field theory with $G$ symmetry that
-- when defined on a manifold $Y$ endowed with a $G$-bundle -- has $\exp(-\pi\i\eta_R^Y/2)$ as its partition function.

To apply this theorem to our problem, we have to be careful with a minus sign.  We start with two manifolds $X$ and $X'$ that each have the same
boundary $W$ (and an extension of the same gauge bundle) as  in fig. \ref{first}(a).  When we say that $X$ and $X'$ have the same boundary $W$,
this is in a convention in which they are on the ``same side'' of $W$, meaning (if they are orientable) that they are oriented in the same way, and in general
in the context of our problem that their $\pin^+$ structures restrict to the same $\pin^+$ structure on $W$.  To glue them to make a manifold $X^*$ without
boundary, we first ``flip over'' $X'$, which means that we reverse its orientation if it is orientable, and in general we replace its $\pin^+$ structure by the complementary one.\footnote{This
notion is described in \cite{Witten}.  If $P$ is a $\pin^+$ structure over $X$, then there is a complementary $\pin^+$ structure $P'$ such that the monodromies
of $P'$ around any orientation-reversing loop in $X$ are the negatives of those of $P$.  ``Flipping'' $X'$ exchanges $P$ and $P'$, and this maps
$\exp(-\pi\i\eta_R^{X'}/2)$ to its inverse, as stated in the text.
To understand why the flip exchanges $P$ and $P'$, let $\psi$ be a fermi field on $X$ and $\psi'$ a fermi field on $X'$.  In gluing $X$ to $X'$ along their common boundary $W$, one sets 
$\psi'|_W=\slashed{n}\psi|_W$, where $\slashed{n}=\vec\gamma\cdot \vec n$, $\vec n$ being the unit normal vector to $W$ in $X$ or $X'$ and $\vec\gamma$ the Dirac gamma matrices.  
Since $\slashed{n}$ anticommutes 
with the gamma matrices of $W$, acting with $\slashed{n}$ reverses the $\pin^+$ structure of $W$.  (A monodromy around an orientation-reversing 
loop in $W$ is represented by a gamma matrix of $W$, so acting with $\slashed{n}$ changes the sign of this monodromy.)   Thus for the gluing relation
$\psi'|_W=\slashed{n}\psi|_W$ to make sense,  $X$ and $X'$ must have opposite $\pin^+$ structures along $W$.  Reversing the $\pin^+$ structure
can be accomplished by changing the sign of all gamma matrices.  This changes the sign of the Dirac operator, so it 
changes the sign of the eta-invariant, and replaces $\exp(-\pi\i\eta_R^{X'}/2)$ with its inverse.}  
Let us write $-X'$ for the flipped version of $X'$.  The gluing involves joining $X$ to $-X'$, as in fig. \ref{first}(b).  ``Flipping'' $X'$ reverses the sign of its eta-invariant.
So in this context, the gluing formula reads
\be\label{glutth}\frac{\exp\left(-\pi\i\eta_R^X/2\right)}{\exp\left(-\pi\i \eta_R^{X'}/2\right)} =\exp(-\i \pi \eta_R^{X^*}/2). \ee  
Clearly, then, 
the left hand side of eqn. (\ref{glutth}) 
is always 1 if and only if the right hand side is always 1.  
So $Z_\Psi$ as defined in eqn. (\ref{torof})  is independent of the choice of $X$ if and only if $\exp(-\pi\i\eta_R^{X^*}/2)$
is 1 whenever $X^*$ has no boundary.  

The physical interpretation is straightforward:  the boundary fermions make a consistent 
$2+1$-dimensional theory by themselves if and only if the bulk topological field theory with partition function $\exp(-\pi\i\eta^X_R/2)$ is
trivial.

\subsection{A More Precise Question, And A Strategy To Answer It}\label{morep}

The question that we have stated so far is a little too crude for some of our applications.
  Even if the emergent gauge group $G$ is absent, the theory of $n$ Majorana
fermions in $2+1$ dimensions (all transforming under $\sT$ with a + sign) is anomalous unless $n$ is a multiple of 16.  In this
paper, we are not mainly interested in this purely gravitational anomaly, which is familiar from several points of view and has been
studied from the present point of view in \cite{Witten} (following a calculation in \cite{R} in a particular example).  
  
We can always reduce to the case that $n$ is a multiple of 16 by adding gauge singlet fermions.  This will eliminate the purely gravitational
anomaly and focus attention on the dependence of the anomaly on the gauge bundle.  Instead of doing this, let us just agree to refine
the question we ask.  Instead of asking whether $\exp(-\pi\i\eta_R/2)$ is always 1 (on a four-manifold $X$ without boundary), we will
ask whether it has a nontrivial dependence on the choice of a $G$-bundle over $X$.  A ``no'' answer to this question
means that the anomaly is purely gravitational, and does not depend on the gauge fields at all.  

In our condensed matter application, we will want to know when it is possible to couple boundary fermions to emergent gauge fields without
any new contribution to the anomaly.  For this, we will need to know precisely when $\exp(-\pi\i\eta_R/2)$ does not depend on the $G$-bundle.
Knowing how to analyze this question will also provide the starting point for our application to string/M-theory.

For our purposes, a very simple criterion will be sufficient.  Let $V_R\to X$  be the real vector bundle associated to the representation
$R$ of $X$.  (Thus, $R$-valued fermions are sections of $P\otimes V_R$, where $P$ is the $\pin^+$ bundle of $X$.  Mathematically, $V_R$
is called the vector bundle associated to the underlying $G$-bundle over $X$ in the representation $R$.)  Since $\exp(-\pi\i\eta_R/2)$
can depend on a choice of gauge bundle only via the topology of $V_R$, if $V_R$ is always trivial then $\exp(-\pi\i\eta_R/2)$ is certainly
independent of the choice of $G$-bundle.

We can generalize this criterion slightly. Let $\mo$ be be a trivial real line bundle over $X$, and let $\vepsk$ be a trivial rank
$k$-bundle constructed as the direct sum of $k$ copies of $\mo$.  It can happen that $V_R$ is not always trivial but there is some
$k$ such that $V_R\oplus \vepsk$ is always trivial.  This means that the field $\Psi$, when supplemented with $k$ 
neutral Majorana fermions (coupled to gravity only), has only a gravitational anomaly.  But the neutral fermions that we had to add
to get this result do not couple to the gauge fields at all, so they only contributed to the purely gravitational part of the anomaly.  Hence
 also
under these more general conditions, the anomaly does not depend on the gauge bundle.

As a matter of terminology, if 
$V_R\oplus \vepsk$ is trivial for some $k$, we say that $V_R$ is stably trivial.  Thus  a fermion system such that $V_R$
is always   stably trivial has only a gravitational anomaly.\footnote{\label{stable} In general, two vector bundles $V$ and $V'$ over  $X$
are said to be stably equivalent if $V\oplus \o^k$ is equivalent to $V'\oplus \o^k$ for some $k$.  Thus for $V$ to be stably trivial means that it is stably
equivalent to a trivial bundle.  If  $V$ and $V'$ are stably equivalent and have a rank  greater than the dimension of $X$, then they are 
actually equivalent. (This can be proved using a triangulation of $X$; $V$ and $V'$ are certainly isomorphic on the 0-skeleton of a triangulation
and one proves inductively that an isomorphism over the $r$-skeleton can be exended over the $r+1$-skeleton.)
 We will not need to know this in our analysis of the
anomaly, but it will be useful later in the paper.}

We will use this criterion in section \ref{examples} 
to show that some fermion systems in $2+1$ dimensions have only a gravitational anomaly -- and in particular become
entirely anomaly-free if a suitable number of gauge-singlet fermions is added to cancel the gravitational anomaly. These are the models
that we will use in section \ref{gb} to construct gapped boundary states of a topological superconductor.  (The application to string/M-theory in section
\ref{mtheory} involves similar but slightly different considerations.) All models we study in 
which $V_R$ is not stably trivial have the opposite property: for these models, the anomaly does depend on the gauge bundle.

\subsection{Some Additional Facts}\label{addf}

A few more facts will greatly simplify the study of examples in section \ref{examples}.  As a preliminary, we ask the following question.
In what sense does the anomaly that we have described involving the eta-invariant generalize the more familiar ``parity'' anomaly
that can be seen on an orientable manifold?

This question may be answered as follows  (see sections 2 and 4 of \cite{Witten} for more detail). 
If $X$ is an orientable four-manifold, and $\D_R$ is the Dirac operator acting on $R$-valued spinors on $X$,
then\footnote{\label{brief} In brief, on an orientable manifold, the spectrum of $\D_R$ has
 a symmetry $\lambda\leftrightarrow-\lambda$ that follows from considerations of chirality, and ensures that the nonzero modes make no net contribution
 to $\eta_R$. Let  $n_+$ and $n_-$ be the numbers of zero-modes of $\D_R$ of positive or negative chirality;  $n_+$ and $n_-$ are both
 even because of a version of Kramers doubling.  Moreover, $\eta_R=n_++n_-$, while $\I=n_+-n_-$, so $\eta_R$ is congruent to $\I$ mod 4.} the  eta-invariant $\eta_R$ of $\D_R$
receives contributions only from zero-modes of $\D_R$, and is congruent mod 4 to the index $\I$ of $\D_R$. 
  This means that $\exp(-\pi\i\eta/2)$
can be replaced by $\exp(-\pi\i\I/2)$.  Since $\I$ is an even integer,
 $\exp(-\pi\i\I/2)$ can be written in a manifestly real (and thus $\sT$-invariant) form as $(-1)^{\I/2}$.   The anomaly is thus a mod 2
question, involving evenness or oddness of the integer $\I/2$, just as  in more traditional treatments of the ``parity'' anomaly.  
The Atiyah-Singer index theorem can be used to write the integer $\I/2$ as a linear combination of gauge and gravitational instanton numbers.
The anomaly $(-1)^{\I/2}$ depends on the gauge bundle if and only if (on a four-manifold $X$ without boundary) the gauge theory instanton
number can make an odd contribution to $\I/2$.  Now let us compare this
to the usual ``parity'' anomaly in three dimensions.  This anomaly can be computed in terms of the spectral
flow of a one-parameter family of three-dimensional 
Dirac operators, and that spectral flow is equal\footnote{See section 2 of \cite{Witten}.} to $\I/2$ on a four-manifold (the mapping torus) constructed from the given one-parameter
family.  So there is a gauge theory contribution to the ``parity'' anomaly in three dimensions if and only if there is a gauge theory contribution
to $(-1)^{\I/2}$ in four dimensions.  

So we can only get something  beyond the standard ``parity'' anomaly if $X$ is unorientable.  This actually leads to a useful
simplification.  A real vector bundle $V_R$ over an unorientable four-manifold $X$ is stably trivial if and only if its Stieffel-Whitney classes vanish:
\be\label{zork}w_1(V_R)=w_2(V_R)=w_4(V_R)=0. \ee  (We need not consider $w_3$, since it vanishes whenever $w_1=w_2=0$.)
Necessity of this criterion is clear, since the Stieffel-Whitney classes are stable characteristic classes of a real vector bundle.  Sufficiency is proved
as follows starting with a triangulation of $X$.  Vanishing of $w_1(V_R)$ ensures that $V_R$ is stably trivial on the 1-skeleton of $X$;
vanishing of $w_2(V_R)$ ensures that it is stably trivial on the 2-skeleton and (since $w_3(V_R)=0$) the 3-skeleton; and finally, vanishing of
$w_4(V_R)$ ensures that $V_R$ is stably trivial on the 4-skeleton and thus on all of $X$.  In the last step, we use the fact that $X$ is unorientable.
If $X$ were orientable, then the integer-valued instanton number of $V_R$ would be an obstruction to stable triviality of $V_R$ on
the 4-skeleton of $X$.  But with $X$ unorientable, which ensures that $H^4(X,\Z)=\Z_2$, the instanton number of a real vector bundle $V_R$ is
only a $\Z_2$-valued invariant, and can be measured by $w_4(V_R)$.  (Instanton number is only a $\Z_2$-valued invariant on an unorientable
manifold, because an instanton, when transported around an orientation-reversing loop, comes back as an anti-instanton.)

\subsection{Examples}\label{examples}

For our first example, we take $G=\U(1)$ or equivalently $\SO(2)$.

Let $R^{(c)}_n$  be a complex one-dimensional representation of $\U(1)$ with charge $n$.
A one-dimensional complex vector space can be viewed as a two-dimensional real
vector space.  So $R^{(c)}_n$ can be regarded as a real two-dimensional representation of $\SO(2)$; when we do this,  we denote it as $R_n$.

A $\U(1)$ gauge field on a four-manifold $X$ is a connection on a complex line bundle $\L$ over $X$.
A field of charge $n$ is a section of $\L^n$.  $\L^n$ can be regarded as a rank two real vector bundle $V_n$ over $X$.
$V_n$ is the real vector bundle associated to the representation $R_n$ of $\SO(2)$.  Now let us specialize to the case that $X$ is unorientable
and ask if $V_n$ is stably trivial.  As explained in section \ref{addf}, this is so if and only if the Stieffel-Whitney classes of $V_n$ vanish.

In general, if $\mathcal U$ is any complex line bundle, and $V$ is the corresponding rank 2 real vector bundle, then the only nonzero Stieffel-Whitney
class\footnote{Apart from the trivial class $w_0(V)=1$.} of $V$ is the mod 2 reduction of $c_1(\mathcal U)$, the first Chern class of $\mathcal U$:
\begin{align}\label{zorf} w_1(V)&=0=w_i(V),~i>2\cr
                                      w_2(V)&= c_1(\mathcal U)~\mod ~2.\end{align} 
(Here $w_i(V)=0$ for $i>2$ since in general the Stieffel-Whitney classes $w_s(V)$ of a rank $r$ real vector bundle $V$
vanish for $s>r$.)   In our case, this means  that the only nonzero Stieffel-Whitney class of $V_n$ is 
\be\label{orf} w_2(V_n)=c_1(\L^n)=nc_1(\L)~\mod~2.\ee
If $n$ is even, then since the Stieffel-Whitney classes are mod 2 classes, this result implies that $w_2(V_n)=0$.
Hence $V_n$ is stably trivial in general (that is, on any four-manifold $X$ and for any gauge bundle) if $n$ is even.   
For odd $n$, this is not true, since in general $nc_1(\L)$ does not vanish mod 2. 

From this we learn that a $\U(1)$ or $\SO(2)$ gauge theory in $2+1$ dimensions with a single fermion multiplet of even charge $n$ is completely 
anomaly-free, even on an unorientable manifold.  For odd $n$, this is not true; in fact, for odd $n$, this theory possesses the standard
parity anomaly.  (The ``parity'' anomaly is most often considered for charge 1, but it is a mod 2 effect and only depends on the charge mod 2;
see section 3.6 of \cite{SW} for a detailed explanation.)

More generally, let us consider a theory with gauge group $\U(1)$ or $\SO(2)$ and with Dirac fermions $\chi_1,\dots,\chi_k$ of charges $n_1,\dots,n_k$.
Here the $n_i$ are integers that (after possibly replacing some of the $\chi$'s with their charge conjugates)
we can take to be nonnegative.  The corresponding real representation of $\SO(2)$ is $R=\oplus_{i=1}^k R_{n_i}$,
and the corresponding real vector bundle is $V_R=\oplus_{i=1}^k V_{n_i}$.
To compute the Stieffel-Whitney classes of $V_R$, we need the Whitney sum formula.  If $V$ is a real vector bundle over a $d$-dimensional
manifold $X$, the total Stieffel-Whitney class of $V$ is defined as $w(V)=1+w_1(V)+w_2(V)+\dots+ w_d(V)$.
The Whitney sum formula says that if $V$ and $V'$ are two such bundles, then
\be\label{sumf}w(V\oplus V')=w(V)w(V'). \ee
This determines the Stieffel-Whitney classes of the direct sum $V\oplus V'$ in terms of those of $V$ and $V'$.  Repeated application of the
formula determines the Stieffel-Whitney classes of a direct sum with more than two summands, such as  $V_R=\oplus_{i=1}^k V_{n_i}$.
We find
\begin{align}\label{wumf} w_2(V_R)& = \sum_i n_i c_1(\L)~\mod ~ 2 \cr
                          w_4(V_R)&= \sum_{i<j}n_in_j c_1(\L)^2~\mod ~ 2.\end{align}

In general, neither $c_1(\L)$ nor $c_1(\L)^2$ vanishes mod 2.   So the condition for $V_R$ to be stably trivial is
\begin{align}\label{umf} \sum_{i=1}^k n_i &=0~\mod~2 \cr
                      \sum_{1\leq i<j\leq k} n_in_j&=0~\mod~ 2.\end{align}                   
Since these are mod 2 conditions, they receive contributions only from those $n_i$ that are odd.  Let $y$ be the number of odd $n_i$'s.
The first condition in (\ref{umf}) says that $y$ is even, and the second says that $y(y-1)/2$ is even.  Taken together, these conditions
say that $y$ is divisible by 4:
\be\label{lumf}y=0~\mod~4.\ee
$V_R$ is stably trivial if and only if this is true.

That $y$ should be even to avoid a gauge anomaly 
is what one would expect from the standard ``parity'' anomaly.  The requirement that $y$ should be divisible by 4
is a stronger condition that goes beyond what one finds on an orientable manifold.  The above argument shows only that a $\U(1)$ or $\SO(2)$
gauge theory is completely anomaly-free if $y$ is divisible by 4.  However, the converse -- there is an anomaly if $y$ is not divisible by 4 --
can be shown by an explicit example that is described momentarily.  
Thus in trying to define a $\U(1)$ gauge theory on an unorientable manifold, we really do find a condition stronger
than the standard ``parity'' anomaly:  the number of fermion multiplets of odd charge must be not just even but divisible by 4.

To demonstrate an anomaly when $y$ is not divisible by 4, 
we should find a four-manifold $X$ and a $\U(1)$-bundle $\L$ over $X$ such that, for a representation
$R$ with $y$ congruent to 2 mod 4, $\exp(-\pi\i\eta_R/2)$ is different from what it would be if $\L$ is replaced by a trivial line bundle.
For this, we take $X=\RP^4$  (see Appendix C of \cite{Witten} for background to the following).  
On $X$, there are two $\pin^+$ structures, say $P$ and $P'$.  They can be distinguished by the eta-invariant of the Dirac operator (coupled
to gravity only), which satisfies $\exp(-\pi\i\eta/2)=\exp(\pm 2\pi\i/16)$, where the sign depends on the $\pin^+$ structure; we choose $P$ to
be the $\pin^+$ structure with $\exp(-\pi\i\eta/2)=\exp(- 2\pi\i/16)$.
Since $\pi_1(\RP^4)=\Z_2$, there is over $\RP^4$  a nontrivial real line bundle $\veps$, and the relation between 
$P'$ and $P$ is $P'=P\otimes \veps$.   
  The Dirac operator $\D_\veps$ for a Majorana fermion, still with $\pin^+$ structure $P$, but coupled
to $\veps$, is the same as the Dirac operator with $\pin^+$ structure $P'$, but coupled to gravity only.  So, writing $\eta_\veps$ for the 
eta-invariant of $\D_\veps$, it satisfies
\be\label{zelkin}\exp(-\pi\i\eta_\veps/2)=\exp(+2\pi\i/16). \ee

If we take the gauge bundle $\L$ to be trivial, then $V_R$ is a trivial bundle of rank $2k$ and hence, for $\pin^+$ structure $P$,
$\exp(-\pi\i \eta_R)=\exp(-2k\cdot 2\pi \i/16)$.  Now, again using the fact that $\pi_1(\RP^4)=\Z_2$,   take $\L$ to be a flat but nontrivial
complex line
bundle over $\RP^4$.  Then $\L^2$ is trivial and so $\L^n$ is trivial or isomorphic to $\L$ for even or odd $n$.  Likewise, the corresponding
rank 2 real vector bundle $V_n$ is a trivial bundle of rank 2 or a copy of $\veps\oplus\veps$ depending on whether $n$ is even or odd.
So $V_R=\oplus_{i=1}^k V_{n_i}$ is the direct sum of a trivial real bundle of rank $2k-2y$ and $2y$ copies of $\veps$.   Now using (\ref{zelkin}),
for the gauge bundle $\L$
  we get $\exp(-\pi\i\eta_R/2)=\exp(-(2k-4y)2\pi\i/16)$.   The condition for this to coincide with the result $\exp(-2k\cdot 2\pi \i/16)$
  that we get when $\L$ is trivial
is simply that $y$ should be divisble by 4.   So there is indeed an anomaly when that condition is not satisfied.  

In section \ref{gb}, we will also want to know what happens for $G=\Z_2$.
In this case, there is only one nontrivial irreducible representation -- a one-dimensional real representation $R_0$ on which the nontrivial element
of $\Z_2$ acts as $-1$.  A $\Z_2$-bundle over a four-manifold $X$ associates to $R_0$ a real line bundle that we will call $\upalpha$.
Now let the representation $R$ consist of the direct sum of $k$ copies of $R_0$.  The corresponding vector bundle $V_R$
is the direct sum of $k$ copies of $\upalpha$.  Its total Stieffel-Whitney class is
\begin{align}\label{omi}w(V_R)=&w(\upalpha)^k=(1+w_1(\upalpha))^k=1+
kw_1(\upalpha)+\frac{k(k-1)}{2}w_1(\upalpha)^2\cr &+\frac{k(k-1)(k-2)}{3!}
w_1(\upalpha)^3+\frac{k(k-1)(k-2)(k-3)}{4!}w_1(\upalpha)^4.\end{align}
This equals 1 for all $X$ and all $\upalpha$ if and only if the coefficients of positive powers of $w_1(\upalpha)$ are all even.
This is so precisely if $k$ is a multiple of 8.  We thus learn that $V_R$ is stably trivial if and only if $k$ is a multiple of 8.  If $k$ is a multiple
of 8, it follows that  the theory has no gauge anomaly.  If $k$ is not a multiple of 8, there can be a gauge anomaly, as one can see again from
the example of $X=\RP^4$.  In this example, with $P$ as before,
$\exp(-\pi\i\eta_R/2)$ is equal to $ \exp(-2k\pi \i/16)$ if $\upalpha$ is trivial, but it equals $\exp(+2k\pi\i/16)$
if $\upalpha=\veps$.   These are only equal if $k$ is a multiple of 8.

In section \ref{confinement}, we will need further examples of the following sort.  Let $G$ be any subgroup of $\SO(p)$, for some $p$.  We view the vector representation $\mathbf p$ of $\SO(p)$
as a real representation $R_0$ of $G$.   Because $G$ is a subgroup of the connected group $\SO(p)$ (rather than ${\mathrm O}(p)$), the
 gauge bundle $V_{R_0}$ associated to this representation satisfies
\be\label{dref}w_1(V_{R_0})=0.\ee
Its higher Stieffel-Whitney classes may not vanish, and Majorana fermions valued in the representation $R_0$ may be  anomalous.
However, let $R=R_0\oplus R_0\oplus R_0\oplus R_0$ be the direct sum of four copies of $R_0$.  The corresponding real vector
bundle is likewise a fourfold direct sum: $V_R=V_{R_0}\oplus V_{R_0}\oplus V_{R_0}\oplus V_{R_0}$.  Using (\ref{dref}) and the Whitney sum
formula, one finds that the Stieffel-Whitney classes of $V_R$ vanish in four dimensions.   Accordingly, $V_R$ is always stably trivial
and  $2+1$-dimensional fermions valued in the representation $R$ have no gauge anomaly.

In section \ref{confinement}, we will make use of two cases of this construction.  For the first case, we take $G=\U(2)$,
and we take $R_0$ to be the two-dimensional complex representation of $\U(2)$, viewed as a four-dimensional real representation.
For the second case, we take $G$ to be the maximal torus $\U(1)\times \U(1)$ of $\U(2)$, with the same $R_0$.  

In either of these examples, if we take only two copies of $R_0$, we would have an anomaly.  In fact, we would have an anomaly
even if we restrict the gauge group to the $\U(1)$ subgroup of $\U(2)$ (or of its maximal torus) embedded as
\be\label{zorff}\begin{pmatrix}*&0\cr 0&1\end{pmatrix}.\ee
Under this subgroup, Majorana fermions valued in $R_0$ are equivalent to a pair of complex fermions of respective charges 1 and 0.
So it we take two copies of $R_0$, the charges are $1,1,0$, and 0.  This is a $\U(1)$ theory with $y=2$, and it is anomalous, as we have seen.

\subsection{The Partition Function}\label{interpretation} 

What we have found can potentially be applied in condensed matter physics in two ways.

If $G$ and $R$ are such that $\exp(-\pi\i\eta_R/2)$ depends on the gauge bundle in a nontrivial way, then it is the partition
function of a four-dimensional topological field theory with $G$ symmetry (defined on $\pin^+$ manifolds and coupled to a background gauge
field).  If this theory is formulated on a manifold $X$ with boundary $W$, then it has a boundary state in which Majorana fermions in the
representation $R$ propagate on $W$.  The partition function of the combined system consisting of the boundary fermions and the 
bulk topological field theory is, as we have explained,
\be\label{bogus} Z_X=|\Pf(\D)|\exp(-\pi\i\eta^X_R/2). \ee
(For a reason that will be clear in  a  moment, we here denote the eta-invariant as $\eta_R^X$ and not just as $\eta_R$.)

If instead $\exp(-\pi\i\eta^X_R/2)$ is always independent of the gauge bundle, then a system of Majorana fermions in $2+1$ dimensions,
transforming in the representation $R$ of $G$ (and transforming with a + sign under time-reversal) has only the usual gravitational 
anomaly that the same
number of neutral fermions would have. It is a conceivable gapless boundary state of a topological superconductor (with $\nu=\mathrm{dim}\,R$, the
real dimension
of $R$), and can be the starting point in
constructing gapped symmetry-preserving boundary states, as we will discuss in section \ref{gb}.  If the dimension of $R$ is 
divisible by 16, the gravitational anomaly is also absent, and 
this system is a completely consistent and anomaly-free $2+1$-dimensional field theory, even on unorientable manifolds.

If the gauge bundle on $W$ can be extended over $X$, then the same formula (\ref{bogus}) serves as the partition function for the combined
system consisting of a bulk topological superconductor and the boundary fermions.  The fact that $\exp(-\pi\i\eta^X_R/2)$ is independent of the
gauge bundle on a manifold without boundary means, by virtue of the gluing formula (\ref{glutth}), 
that the choice of how the gauge bundle is extended over $X$ does
not matter.

If the gauge bundle on $W$ cannot be extended over $X$, how can we describe the partition function of the combined system consisting of the bulk topological superconductor on $X$ coupled
to boundary fermions on $W$?   It is easier to explain
what to do if there is {\it some} four-manifold $X'$ over which the gauge bundle and $\pin^+$ structure of $W$ can be extended, so we begin with this case.
Let $\eta_R^{X'}$ be the eta-invariant of the Dirac operator on $X'$.  For a further simplification, assume first that the dimension of $R$
is divisible by 16, so that the fermion system on $W$ is completely anomaly-free.  Then $X$ plays no role, and since the partition function
depends only on $W$, we denote it as $Z_W$.  For $Z_W$ we can use the same formula as in eqn. (\ref{bogus}), but with $\eta_R^X$ replaced
by $\eta_R^{X'}$ on the right hand side:
\be\label{ogus} Z_W=|\Pf(\D)|\exp(-\pi\i\eta^{X'}_R/2). \ee
The usual arguments show that this does not depend on the choice of the $\pin^+$ manifold $X'$ and the extension of the gauge bundle over $X'$.

What if $\nu=\mathrm{dim}\,R$ is not a multiple of 16?  Then the bulk topological superconductor on $X$ is nontrivial, and (\ref{ogus})
cannot be correct; it describes a topological superconductor with boundary $W$ whose worldvolume is $X'$ rather than $X$!   To fix
the situation, let us recall the four-manifold $X^*$ without boundary that is built by gluing $X$ and $X'$ along their common boundary
(fig. \ref{first}).  Let $\eta_0^{X^*}$ be the eta-invariant of the Dirac operator on $X^*$ coupled to gravity only.  Then the appropriate
generalization of (\ref{ogus}) for the partition function is
\be\label{togus}Z_{W;X}=|\Pf(\D)|\exp(-\pi\i\eta^{X'}_R/2)\exp(-\nu\pi\i\eta_0^{X^*}/2). \ee
We denote this partition function as $Z_{W;X}$ because -- since the bulk topological superconductor on $X$ is now nontrivial -- it does depend on
$X$, not just on $W$.  The gluing theorem for the eta-invariant can be used to show that this
 formula  does not depend on the choice of the $\pin^+$ manifold $X'$ and the extension over $X'$
of the gauge bundle and $\pin^+$ structure.  (Eqn. (\ref{togus})  is analogous to eqn. (3.41) in \cite{SW}, where $X$ was assumed to be orientable
and a coupling to electromagnetism was incorporated.)

However, it may happen that a suitable $X'$ does not exist, because $W$ with its $G$-bundle may represent a nontrivial element of the
group $\Gamma$ that classifies up to cobordism three-dimensional $\pin^+$ manifolds $W$ endowed with a $G$-bundle.  
Note that $W$ is always trivial in cobordism if we ignore the $G$-bundle (since it is given as the boundary of the worldvolume $X$ of the
topological superconductor) but it may be that $W$ together with its $G$-bundle is not a boundary.\footnote{An obstruction is given by the mod 2
index of the Dirac operator on $W$ with values in any real representation of $G$.} 
The
physical meaning of $\Gamma$ is that \cite{FH}  $\mathrm{Hom}(\Gamma,\U(1))$ parametrizes a family of invertible three-dimensional
topological field theories defined
on a $\pin^+$ manifold $W$ with a $G$-bundle.  Concretely, suppose we are given a homomorphism $\varphi:\Gamma\to \U(1)$.
The associated topological field theory $\T$ may be described as follows.  Let $W$ be a three-manifold with $\pin^+$ structure and a $G$-bundle.
Then $W$ has a class $[W]$ in $\Gamma$ and the partition function of the theory $\T$ on $W$ is $\varphi([W])$.    This is relevant because when such theories exist, we should not expect to get
a unique answer for the theory of fermions on $W$ in the representation $R$ of $G$ coupled to the topological superconductor on $X$.
We could always modify any possible answer by tensoring by a three-dimensional topological field theory that depends only on the boundary
data.

Actually, since in our application $W$ is a boundary if we forget the $G$-bundle, $\Gamma$
is not quite the right group to consider.  We should replace $\Gamma$ by its  subgroup $\Gamma'$ that classifies up to cobordism
pairs consisting of a $\pin^+$ manifold $W$ with a $G$-bundle, such that $W$ is a boundary if we forget the $G$-bundle.
To explain how to proceed, suppose for example that $\Gamma'=\Z_2$.  Let $W_0$, together with some $G$-bundle, generate
$\Gamma'$. There is then a three-dimensional topological field theory $\T$,
defined on $\pin^+$ manifolds $W$ with $G$-bundle, such that $W$ is a boundary if we forget the $G$-bundle,
 that assigns the value $-1$ to $W_0$ and is trivial if $W$ with its $G$-bundle is a boundary.

Pick a $\pin^+$ manifold $X_0$ with boundary $W_0$.  (By hypothesis, the $G$-bundle over $W_0$ does not extend over $X_0$.)
In this situation, we should not expect to get a unique answer for
$Z_{W_0;X_0}$, because any answer could be modified by tensoring with $\T$.  However, let $X_0\sqcup X_0$ be the disjoint union of
two copies of $X_0$, with boundary $W_0\sqcup W_0$.   Then $W_0\sqcup W_0$, together with its gauge bundle, is a boundary
(we have assumed that $\Gamma'=\Z_2$ so two copies of $W_0$ make a boundary), so the formula (\ref{togus}) can be used to determine the partition function $Z_{W_0\sqcup W_0;X_0\sqcup X_0}$.
We interpret this partition function as  $Z_{W_0;X_0}^2$ (the square of $Z_{W_0;X_0}$)
and we make some choice of the square root to determine  $Z_{W_0;X_0}$.   Now let $W$ be any three-dimensional $\pin^+$
manifold with a $G$-bundle.  Then since $W_0$ generates $\Gamma'$,
either $W$ or $W\sqcup W_0$ is (together with its $G$-bundle) the boundary of some $X'$.  Hence
eqn. (\ref{togus}) can be used to determine either $Z_{W;X}$ or $Z_{W\sqcup W_0;X\sqcup X_0}$, which we 
 interpret as the product $Z_{W;X} Z_{W_0;X_0}$.
$Z_{W_0;X_0}$ has already been determined, so in any case we arrive at a result for $Z_{W;X}$.  Thus, this gives a complete description of the
partition function in general.

  In this construction, we made an
arbitrary choice of sign of $Z_{W_0;X_0}$. 
 The two possible
choices of sign will give results that differ by tensoring with the purely $2+1$-dimensional
invertible topological field theory $\T$.   We also assumed that $\Gamma'=\Z_2$.
In general, $\Gamma'$ is always a finite abelian group.  The above reasoning can be modified in a fairly obvious way for any $\Gamma'$.

\section{Gapped $\sT$-Invariant Boundary States Of A Topological Superconductor}\label{gb}

In this section, we describe gapped $\sT$-invariant boundary states of a topological superconductor 
with even $\nu$. (The methods we use do not suffice
to construct such states for odd $\nu$.)   The class of models we consider and most of the methods for analyzing them were already presented
in \cite{SW} (boundary states for topological superconductors are treated in section 6 of that paper, following a similar treatment of topological
insulators earlier in the paper).  We can add something here because of having a more complete knowledge of which models are anomaly-free.
However, we will be brief because many points were already explained in \cite{SW}.   See also related analysis in \cite{WS,MCFV,MKF,TPfaffian,TPfaffiantwo,Wangetal,Wangetals,MV,Al,VS,MetlitskiTSC}.

We begin with an overview and then describe some properties of models for different values of $\nu$.
As in \cite{SW}, it turns out that the main results for all values of $\nu$ can be described using considerations of weak coupling.  

\subsection{Overview}\label{overview}

We consider a bulk topological superconductor with a given even $\nu$.  The standard boundary state would consist of $\nu$ massless
Majorana fermions, all transforming with a $+$ sign under $\sT$.  As before, we denote the worldvolume of the topological superconductor as $X$ and
its boundary as $W$.  

To construct a new type of boundary state, we postulate the appearance on $W$ of an emergent gauge symmetry with gauge group $G$.
We also assume that on $W$ there propagate $\nu$ Majorana fermions, transforming with a $+$ sign under $\sT$ and in some $\nu$-dimensional real
representation $R$ of $G$.  The representation is chosen to have no gauge anomaly according to a criterion explained in section \ref{analysis}.
Thus the only anomaly is the standard gravitational anomaly of $\nu$ Majorana fermions that transform the same way under $\sT$.
This is then a conceivable (gapless) boundary state of a topological superconductor with the given value of $\nu$.

More generally, we could pick a positive integer $s$, and a pair of real representations $R$ and $R'$ of $G$, of respective dimensions $\nu+s$ and $s$.
Then we introduce Majorana fermions $\Psi$ and $\Psi'$ transforming respectively with a $+$ sign or a $-$ sign under $\sT$, and respectively in the
representations $R$ and $R'$ of $G$.  If the anomalies cancel between\footnote{According to the analysis in section \ref{analysis},
this cancellation means that if $X$ is a four-dimensional $\pin^+$ manifold without boundary, 
and $\eta_R$ and $\eta_{R'}$ are the eta-invariants of the
$R$- or $R'$-valued Dirac operator on $X$, then $\exp(-\pi\i (\eta_R-\eta_{R'})/2)$ is indendent of the choice of $G$-bundle over $X$.} $R$ and $R'$, then this gives another conceivable gapless
boundary state of the same topological superconductor.  In the absence of the gauge symmetry, pairs of Majorana fermions transforming oppositely
under $\sT$ can combine and get a mass, but $R$ and $R'$ can be chosen so that the gauge symmetry prevents this.

Now we introduce two sets of charged scalar fields $w$ and $\phi$, chosen to make possible two phases with the following properties.

In one phase, the fields $w$ have expectation values that completely break the emergent gauge symmetry while preserving $\sT$ symmetry.  For example, if $G=\U(1)$,
$w$ can consist of a single complex scalar field of charge 1, with a suitable transformation law under $\sT$.   In the phase in which $w$ has an expectation value,
the emergent gauge symmetry is completely broken and disappears at low energy.  Since we also assume that $\phi$ is massive in this phase, it follows that in this
phase, the only gapless modes are $\nu$ Majorana fermions transforming with a $+$ sign under $\sT$ (or $\nu+s$ transforming with a $+$
sign and $s$ transforming with a $-$ sign, for some $s$; once the gauge symmetry is completely broken, such extra modes can combine
and get masses).  Thus this phase reduces at low energies to the standard boundary state of a topological superconductor.

The second phase, in which $\phi$ gets an expectation value, is supposed to be gapped and $\sT$-conserving.
Assuming that the gap should be visible semiclassically, this means that the expectation value of $\phi$  has to break the gauge symmetry down
to a finite subgroup.  An unbroken subgroup of positive dimension would lead to a gapless spectrum in perturbation theory.  The possibility of
an unbroken subgroup of positive dimension that nevertheless leads to a gapped spectrum through nonperturbative effects is the topic
of section \ref{confinement}.  In the present section, we assume that $\langle\phi\rangle$ breaks $G$ down to a finite subgroup $\K$.
It will be clear in a moment that this subgroup cannot be trivial.

To get a gapped state, $\phi$ must also have Yukawa couplings to the fermions, so that the fermions acquire mass from the expectation
value of $\phi$.  A linear Yukawa coupling of scalar fields $\phi_a$, $a=1,\dots,t$ to Majorana fermions $\psi_i$, $i=1,\dots,\nu$
takes the general form
\be\label{zorof} L_\Yuk = \sum_{i,j=1}^\nu \sum_{a=1}^t m_{ij}^a\bar\psi_i\psi_j \phi_a, \ee
for some constants $m_{ij}^a$.  The transformations of $\phi_a$ under $G$ and under $\sT$ and the constants $m_{ij}^a$ must
be chosen so that this coupling (along with possible nonlinear couplings of $\phi$ to the fermions)
 preserves the symmetries and gives mass to all fermions in the phase with $\langle\phi\rangle\not=0$.  
 
One general comment is that  in a phase in
which the fermions get a mass from the expectation value of $\phi$, that expectation value can never be invariant under
the microscopic time-reversal transformation that acts on the fermions as
\be\label{zeb}\sT(\psi(t,\vec x))=\g_0\psi(-t,\vec x),\ee regardless of how this symmetry is assumed to act on $\phi$.
Indeed, any fermion bare mass term is odd under $\sT$ acting as in (\ref{zeb}). 
Since $L_\Yuk$ is by hypothesis $\sT$-invariant, 
 the expectation value of $\phi$ will necessarily be also odd under $\sT$.
 The time-reversal transformation $\slT$ that
is a symmetry of the gapped phase is therefore not the microscopic $\sT$ that acts on the fermions as in (\ref{zeb}),
 but the combination of $\sT$
with a gauge symmetry, which we call $\sK^{1/2}$:
\be\label{zubber} \slT=\sT \sK^{1/2}.\ee
For $\langle\phi\rangle$ to be $\slT$-invariant, given that it is  odd under $\sT$, it must be odd under $\sK^{1/2}$.

From a microscopic point of view, one has $\sT^2=(-1)^\sF$.  Therefore
\be\label{ubber}\slT^2=(-1)^\sF \sK. \ee
Note that $\sK$ is an unbroken symmetry, since it leaves fixed the expectation value of $\phi$ (by contrast, $\sK^{1/2}$ is not an unbroken gauge
symmetry as the expectation value of $\phi$  is odd under $\sK^{1/2}$).
We should note that, since $\sK$ is an element of the group $G$ of emergent gauge symmetries, any state of a compact sample has $\sK=1$
and hence satisfies the standard relation
$\slT^2=(-1)^\sF$, as one would expect microscopically.  But individual quasiparticles can have $\sK\not=1$, and the
relation (\ref{ubber}) is important in understanding their properties.

The definition of $\slT$ is not unique, because in eqn. (\ref{zubber}), we could multiply $\sK^{1/2}$ by any unbroken gauge symmetry
$\sf S$ (that is, any ${\sf S}\in G$ that leaves fixed $\langle\phi\rangle$) without changing the fact that $\slT$ is a symmetry of the phase
with $\langle\phi\rangle\not=0$.   However, it turns out that no matter what choice of $\slT$ we make, $\sK$ is always nontrivial
if the bulk topological superconductor is nontrivial, that is, if $\nu$ is not a multiple of 16.
This is not immediately obvious because, as $\sK$ leaves fixed the expectation value of $\phi$, one can ask if it is possible to have $\sK=1$.
We may argue as follows.  If $\sK=1$, then $\sK^{1/2}$ generates a $\Z_2$ subgroup of the gauge group $G$. 
All fermion fields are therefore either even or odd under $\sK^{1/2}$.  Let $\phi_0$ be the component of $\phi$ that has an expectation value;
thus $\phi_0$ is odd under $\sK^{1/2}$. A $\sK^{1/2}$-invariant Yukawa coupling that involves
$\phi_0$ has to take the form
\be\label{yrf} \psi_{\mathrm e}\psi_{\mathrm o}\phi_0 \ee
where $\psi_{\mathrm e}$ is even under $\sK^{1/2}$ and $\psi_{\mathrm o}$ is odd.  If all fermions are supposed to gain mass in this
way, the number of fermions even under $\sK^{1/2}$ must be the same as the number of fermions that are odd.
However, we have assumed that the full collection of fermions has  no
gauge anomaly under the full $G$ symmetry, so certainly there is no anomaly under the $\Z_2$ subgroup of $G$ generated by $\sK^{1/2}$.
According to the analysis of anomaly-free $\Z_2$ gauge theories in section \ref{examples}, to avoid a gauge anomaly, the number of
fermions odd under $\Z_2$ must be a multiple of 8; allowing for an equal number of $\Z_2$-even fermions, the total number of fermions
is a multiple of 16 and the bulk topological superconductor is trivial.

The fact that the unbroken gauge symmetry $\sK$ is nontrivial (if the bulk topological superconductor is nontrivial) means that the boundary state,
though gapped, is topologically nontrivial.  It supports a group $\K$ of emergent gauge symmetries that at least contains a cyclic
subgroup generated by $\sK$.   

In bulk, a topological superconductor has a symmetry group $\T$ generated by $\sT$; it is isomorphic to $\Z_4$, the four elements being
 $\sT$, $\sT^2=(-1)^\sF$, $\sT^3=\sT(-1)^\sF$, and $\sT^4=1$.  The boundary state that we have described has a group $\K$ 
 of emergent unbroken gauge symmetries.   But (assuming that the bulk superconductor is topologically nontrivial)
 the  full symmetry group is not a product $\H=\T\times \K$.  Rather, it is a nontrivial extension
 \be\label{morg} 1\to \K\to \H\to \T\to 1. \ee
The extension is nontrivial because we have shown above that no matter how we lift $\sT\in \T$ to $\slT\in\H$, we get
$\slT^2=(-1)^\sF\sK$ with $\sK$ a nontrivial element of $\K$.   In other words, there is no way to lift $\T$ to a subgroup of $\H$.

\subsection{Examples Based On $\U(1)$}\label{uone}

\subsubsection{Generalities}\label{generalities}

Following \cite{SW}, to construct concrete examples, we will take $G=\U(1)$ or equivalently $\SO(2)$. We denote the $\U(1)$ gauge field as $a$
and its field strength as $f=\d a$. We consider only the simplest
examples for different values of $\nu$.  It is convenient to work with a basis of Dirac fermions; that is, instead of considering a pair
$\psi_1,\psi_2$ of Majorana fermions transforming in a two-dimensional real representation of $\SO(2)$, we combine them
to a Dirac fermion $\chi=(\psi_1+\i\psi_2)/\sqrt 2$ and its adjoint $\bar\chi$. They have equal and opposite $\U(1)$ charge.
 Reversing the sign of $\psi_2$ would exchange $\chi$ and $\bar\chi$,
so we can assume that $\chi$ has nonnegative $\U(1)$ charge.  

As in \cite{SW}, one does want not the macroscopic theory to have an exact symmetry that has no microscopic origin, so one does not
want $j_\mu=\epsilon_{\mu\nu\lambda}f^{\nu\lambda}/4\pi$ (where $\epsilon_{\mu\nu\lambda}$ is the Levi-Civita
tensor of the boundary) to be a conserved current. This is prevented by adding a monopole operator to the Lagrangian or Hamiltonian.
In all models that we will consider, there is a symmetry-preserving monopole operator of unit magnetic charge that can be added.

For $G=\U(1)$, we can take $w$ to consist of a single complex scalar field of charge 1, whose expectation value would completely break
the gauge theory.  It will also suffice, as in \cite{SW}, to take $\phi$ to consist of a single complex scalar field.    Expanding $w$ and $\phi$
in terms of real scalar fields as $w=(w_1+\i w_2)/\sqrt 2$, $\phi=(\phi_1+\i\phi_2)/\sqrt 2$, we take\footnote{The following transformations
are the transformations under $\sf{CT}$ given in  eqns. (3.3)-(3.6) of \cite{SW}.  What we call $\sT$ in the present paper was called $\sf{CT}$ in \cite{SW},
because that paper focused primarily on topological insulators rather than superconductors.} $\sT$ to act by $\sT(w_i)=w_i$,
$\sT(\phi_i)=-\phi_i$.  We describe this by saying that $w$ is $\sT$-even and $\phi$ is $\sT$-odd.  
The fact that the $w_i$ are $\sT$-even means that the expectation value $\langle w\rangle$ can break the emergent  $\U(1)$ gauge
symmetry completely while conserving $\sT$.  The fact that the $\phi_i$ are $\sT$-odd ensures
$\sT$-invariance of the Yukawa coupling (\ref{zorof}).   

Now suppose that $\phi$ has charge $p$
under $\U(1)$.  Then to ensure that the expectation value of $\phi$ is invariant under $\slT=\sT \sK^{1/2}$, we can take
$\sK^{1/2}$ to be a gauge transformation by $\exp(2\pi\i/2p)$.  $\sK$ is then a gauge transformation by $\exp(2\pi\i/p)$, and generates
the group $\sK\cong \Z_p$ of unbroken gauge symmetries.

  If $p$ is even and $\chi$ is a Dirac fermion of charge $p/2$, then a Yukawa coupling
\be\label{orbo} \sum_{a,b=1}^2\epsilon^{ab}\chi_a\chi_b \bar\phi+\mathrm{h.c.} \ee
(here $a$ and $b$ are spinor indices) is $\U(1)$-invariant and $\sT$-invariant and gives $\chi$ a mass.  Similarly, if $\chi $ and $\chi'$
are two Dirac fermions with charges $t$ and $p-t$, then they can receive a mass from a coupling
\be\label{norbo}\sum_{a,b=1}^2\epsilon^{ab}\chi_a\chi'_b \bar\phi+\mathrm{h.c.} \ee  We will construct models by combining these ingredients.
It is also possible to introduce pairs of fermions that receive masses from couplings cubic or higher order in $\phi$.

If $\phi$ has charge $p$, then a vortex in which the phase of $\phi$ twists by $2\pi$ has a flux $\int f/2\pi=1/p$.  A monopole
operator shifts the flux by an integer, so after adding a monopole operator of unit charge to the action, the vorticity is conserved mod $p$. 
In fact, conservation of vorticity mod $p$ is a feature of low energy $\Z_p$ gauge theory.

\subsubsection{Monopole Operators}\label{monop}

The assertion that vorticity is conserved precisely mod $p$ depends on existence of a charge 1 monopole operator that is a Kramers
singlet boson of spin 0 and hence can be added to the action or Hamiltonian.  Such an operator exists for any model that lacks the generalized ``parity''
anomaly of section \ref{examples}, but the point is somewhat subtle.   This question has been addressed
in \cite{WS,MCFV}, in a slightly different language, for Dirac fermions of charge 1, and in \cite{SW} for fermions of even charge as well as some other
cases (the latter analysis was made mainly  in the context of a topological insulator rather than superconductor).

We first consider a Dirac fermion $\chi$ of even charge $2s$, which makes no contribution to the anomaly so it can be considered by itself.
We determine the quantum numbers of a charge 1 monopole
operator in a standard way via radial quantization on $S^2$ \cite{K}.   In a sector with $\int_{S^2}f/2\pi=1$, the field $\chi$ 
has $2s$ zero-modes $\zeta_a$, $a=1,\dots,2s$ of spin $(2s-1)/2$ and $\U(1)$ charge $2s$, while its adjoint $\chi^\dagger$ has the same number of zero-modes $\zeta_a^\dagger$
of the same spin and charge $-2s$.  One can choose $4s$ real linear combinations of these modes, but those real linear combinations are not $\sT$-invariant.
That is because of the factor of $\g_0$ in the transformation of a Majorana fermion under time-reversal, which we have taken to be
\be\label{worb}\sT\psi(t,\vec x)=\g_0\psi(-t,\vec x).\ee  However, it is possible to define a modified $\sT$ transformation under which the real zero modes are all invariant.
This is
\be\label{zorb}\slT'=\sT \exp(-2\pi\i\q/8s), \ee
where $\q$ is the $\U(1)$ charge generator.  The basis for this statement is as follows. 
The zero-modes of $\chi$ and $\chi^\dagger$
 have respectively positive and negative chirality along $S^2$. The chirality in the 2d sense is measured by the operator $\i\g_1\g_2$.
In $2+1$ dimensions, one can take the Dirac $\gamma$ matrices to be $2\times 2$
real matrices that obey $\g_0\g_1\g_2=1$, so actually $\i\g_1\g_2=-\i\g_0$.  Thus the $\g_0$ that appears in the transformation law (\ref{worb}) acts as
$\i$ on zero-modes of $\chi$ and as $-\i$ on zero-modes of $\chi^\dagger$.  The factor $\exp(-2\pi\i\q/8s)$ in the definition of $\slT'$ compensates for this 
and ensures that $\slT'$ acts on fermion zero-modes only by complex conjugation, leaving real zero-modes invariant. 

The total number of $\chi$ and $\chi^\dagger$ zero-modes is $4s$.  Quantizing $4s$ real fermion modes gives a Hilbert space $H$ of dimension $2^{2s}$.
An anti-unitary operator $\slT'$ that commutes with all of the real fermion modes acts on this space as 
\be\label{zebb}(\slT')^2=(-1)^s. \ee
This reflects the fact that the spinor representation of $\Spin(4s)$ is pseudoreal or real for odd or even $s$.

The state in $H$ of lowest $\U(1)$ charge is a state $|\negthinspace\downarrow\rangle$ that is annihilated by the $\zeta_i^\dagger$; its $\U(1)$ charge is $-2s^2$ 
(as in a classic analysis \cite{JR}, this follows from $\sT$-invariance, which implies that the state $|\negthinspace\uparrow\rangle =\zeta_1\zeta_2\dots\zeta_{2s}|\negthinspace
\downarrow\rangle$ of highest $\U(1)$ charge must have opposite charge from  $|\negthinspace\downarrow\rangle$).  
The state $|\negthinspace\downarrow\rangle$
is the unique state in $H$ of lowest charge, so it transforms under rotations of $S^2$ with spin zero.  The $\U(1)$-invariant
states in $H$ are obtained by acting $s$ times with the modes $\zeta_a$: they have the form $|a_1a_2\dots a_s\rangle=\zeta_{a_1}\zeta_{a_2}\dots \zeta_{a_s}|\negthinspace
\downarrow\rangle$.  Since the modes $\zeta_a$ have spin $(2s-1)/2$, these states have spin $s/2$ mod 1.  For a $\U(1)$-invariant state, $\slT'$ coincides
with $\sT$, so eqn. (\ref{zebb}) implies that these states obey $\sT^2=(-1)^s$.  If $s$ is even, these states are Kramers singlet bosons, and a real linear
combination of the corresponding monopole operators  can be added to the action or Hamiltonian.  If $s$ is odd, 
the monopole operators $M_{a_1a_2\dots
 a_s}$
corresponding to these states are Kramers doublet fermions.  But we can get a charge 1 monopole operator that is a Kramers singlet boson by multiplying
$M_{a_1a_2\dots a_s}$ by an ordinary gauge-invariant local operator that is a Kramers doublet fermion, such as $\bar w^{2s}\chi$.   So in all cases, there is a
charge 1 monopole operator that is a boson and a Kramers singlet.

This argument shows that one or more Dirac fermions of even $\U(1)$ charge produce no anomaly in the quantization of the monopole operators, in the sense that those operators have the same quantum numbers as ordinary local operators.
So it suffices to consider the case of Dirac fermions $\chi_1,\dots,\chi_y$ that all have odd charges $n_1,\dots,n_y$.  We cannot expect to make the argument in quite the
same way, because individual Dirac fermions with odd charge do contribute anomalies, and we have to combine them to demonstrate
anomaly cancellation.  In imitating the above argument, one's first thought might be to replace $2s$ in the definition of $\slT'$ by the fermion charge $n_i$.  
The trouble with this is that in general $n_i$
depends on $i$.  In demonstrating anomaly cancellation between different $\chi_i$, one really would like to use the same $\slT'$ for all of them.
It turns out that there is a choice that works nicely, given that the $n_i$ are all odd:
\be\label{qebb}\slT'=\sT \exp(-2\pi \i \q/4). \ee
With this choice, the zero-modes of $\chi_i$ are $\slT'$-even if $n_i$ is congruent to 1 mod 4, and $\slT'$-odd if $n_i$ is congruent to 3 mod 4.
So let us rearrange the $n_i$ as integers $u_1,\dots, u_p$ that are congruent to 1 mod 4 and integers $v_1,\dots,v_{p'}$ that are congruent to 3 mod 4.
(Thus $p+p'=y$.)  The total number of real fermion zero-modes that are $\slT'$-even is $a_+=2\sum_i u_i$, and the number that are $\slT'$-odd is 
$a_-=2\sum_j v_j$.  When we quantize all the zero-modes, we get a Hilbert space $H$ of dimension $2^{\sum_i u_i +\sum_j v_j}$.  The state
of lowest $\U(1)$ charge in this Hilbert space is a state $|\negthinspace\downarrow\rangle$ annihilated by all modes of the $\chi_i^\dagger$;
its charge is
\be\label{tormo}q=-\frac{1}{2}\left(\sum_i u_i^2+\sum_j v_j^2\right).\ee
This is an integer if and only if the total number $y$ of Dirac fermions is even.  If $y$ is odd, then $q$ is a half-integer and there is no $\U(1)$-invariant
state in the whole Hilbert space (boson and fermion modes of nonzero energy have integer $\q$ and do not help).  This is a manifestation of the usual ``parity'' anomaly.  Let us assume henceforth that $y$ is even.  This implies
that $a_++a_-$ and $a_+-a_-$ are multiples of 4.
When we quantize $a_+$ $\slT'$-even real fermion modes and $a_-$ $\slT'$-odd ones, with both of these numbers multiples of 4,
 we get a Hilbert space $H$ on which\footnote{One way to see this is to observe that if $\zeta$ and $\zeta'$ are real fermion modes
that transform with opposite signs under $\slT'$, then on the Hilbert space $H$, we can define the $\slT'$-invariant ``Hamiltonian'' $h=\i\zeta\zeta'$.
Reducing to ground states of $h$, we can eliminate $\zeta$ and $\zeta'$ and replace $H$ with a subspace of one-half the dimension on which
$(\slT')^2$ acts with the same sign as before.  Continuing in this way, we reduce to the case that $a_+$ or $a_-$ is 0.  If, say, $a_-=0$, to determine
the sign we must need to know if the spinor representation of $\Spin(a_+)$ is real or pseudoreal, and this leads to eqn. (\ref{wormo}).}
\be\label{wormo} (\slT')^2=(-1)^t,\ee
where
\be\label{frev}t =\frac{a_+-a_-}{4}=\frac{1}{2}\left(\sum_i u_i -\sum_j v_j\right).\ee  
Now we have to find a $\U(1)$-invariant state of monopole charge 1.  For completely general odd integers $n_1,\dots,n_y$, there is no such
state in the Hilbert space $H$ obtained by quantizing the fermion zero-modes.  However, by using the charge 1 scalar field $w$, we can make
the $\U(1)$-invariant state $|\Psi\rangle= w^{-q}|\negthinspace\downarrow\rangle$.  In a field of monopole charge 1, $w$ has spin 1/2, so the
state $|\Psi\rangle$ has spin $q/2$ mod 1.  Since $w$ is a Kramers singlet under $\sT$, and has charge 1, it is a Kramers doublet of $\slT'$.
From eqn. (\ref{wormo}), $(\slT')^2|\negthinspace\downarrow\rangle =(-1)^t|\negthinspace\downarrow\rangle$,  so $(\slT')^2|\Psi\rangle=(-1)^{t-q}|\Psi\rangle$.
But on the $\U(1)$-invariant state $\Psi$, $\slT'$ is the same as $\sT$:
\be\label{zobbo}\sT^2|\Psi\rangle =(-1)^{t-q}|\Psi\rangle. \ee
So $|\Psi\rangle$ is a gauge-invariant state of spin $q/2$ mod 1 that transforms under $\sT^2$ with a sign $(-1)^{t-q}$.  The condition that this state is either
a Kramers singlet boson or a Kramers doublet fermion is that $t$ should be even.    But the formula (\ref{frev}) shows that any of the $\chi_i$
of odd electric charge, whether its charge is congruent to 1 or 3 mod 4, makes a contribution to $t$ that is $1/2$ mod 2.  Hence requiring $t$ to 
be even is the same as requiring that the total number $y$ of Dirac fermions of odd charge is a multiple of 4.  This was the condition found in section
\ref{examples} for the theory to be completely free of gauge anomaly on an unorientable manifold.  

As in our discussion of the even charge case, once we have found a monopole operator $M$ that is either a Kramers singlet boson or a Kramers
doublet fermion, by acting if necessary with a gauge-invariant Kramers doublet fermion $\bar w^{n_i}\chi_i$, we get a monopole operator $M'$
that is a Kramers singlet boson.  So such an operator exists whenever the generalized ``parity'' anomaly is absent.

We turn next to examples for various values of $\nu$.

\subsubsection{$\nu$ Congruent To 2 mod 4}\label{two}

The most minimal model with $\nu=2$ was analyzed in detail in \cite{SW}.  One simply introduces one $\chi$ field with some charge $r$,
and to make a mass term possible, one takes $\phi$ to have charge $2r$.   One chooses $r$ to be an even number $r=2s$ to avoid
the ``parity'' anomaly.  As we know from section \ref{examples}, this gives a theory with no gauge anomaly, not even the more subtle
anomaly that can only be detected on an unorientable manifold.   

Since $\phi$ has charge $4s$, the
 expectation value of $\phi$ reduces the gauge symmetry from $\U(1)$ to $\Z_{4s}$.  The low energy theory, however, is not just a $\Z_{4s}$
gauge theory.  In a vortex field, there is a single Majorana fermion zero-mode.
As a result, the vortices have nonabelian statistics, and, as in the composite fermion description \cite{composite} of the Moore-Read state
\cite{MRe}, the theory can be described, at least on an orientable manifold, as the quotient by $\Z_2$ of a $\Z_{4s}$ gauge theory times
an Ising topological field theory.  For details, see \cite{SW}.    

For any even $\nu=2t$, one simple approach is to introduce $t$ $\chi$ fields, all with the same even charge and coupled in the same way to $\phi$.
This family of models was discussed in section 6.3 of \cite{SW} for odd $t=1,3,5,7$, corresponding to $\nu=2,6,10,14$.  As long as one
is on an orientable manifold, the resulting topological field theories are the same, but they have four different actions of $\sT$ (mainly
because of subtleties in the quantization of the vortices).    On an unorientable manifold, these models are inequivalent at low energies, as is 
clear both from the different values of $\nu$ and from the different actions of $\sT$.   

For $\nu$ congruent to 2 mod 4, the unbroken gauge symmetry is always $\Z_{4s}$ for some integer $s$, even if the $\chi$ fields do not
all have the same charge.  This follows from the constraints
of avoiding the ``parity'' anomaly and ensuring that all fermions gain mass from the gauge symmetry breaking.  (If $\phi$ has charge $4s+2$
for some integer $s$, then, to give masses to all $\chi$ fields, the number of $\chi$ fields with odd charge will have to be odd, leading
to a ``parity'' anomaly.)  In general, if formulated on an orientable manifold,
the $\Z_{4s}$ gauge theory has a Chern-Simons (or Dijkgraaf-Witten \cite{DW}) term, as analyzed in \cite{SW}.

\subsubsection{$\nu$ Divisible By 4}\label{four}

For $\nu=4$, we need two $\chi$ fields.  Naively they can both have odd charge, avoiding the usual ``parity'' anomaly, but we learned
in section \ref{examples} that this will actually lead to a more subtle gauge anomaly.  So we take them to have even charge.

Two Dirac fermion fields $\chi_1,\chi_2$ of charges $2r_1$ and $2r_2$ can gain mass from a Yukawa coupling to a $\phi$ field of charge $2(r_1\pm r_2)$.
(The relevant Yukawa coupling is $\chi_1\chi_2\bar\phi$ or $\chi_1\bar\chi_2\bar\phi$.)  Since $r_1\pm r_2$ can be any integer $r$,  the low energy gauge symmetry can be
$\Z_{2r}$ for any $r$.  

As soon as $\nu$ is a multiple of 4, the number of $\chi$ fields is even and likewise the number of fermion zero-modes in a vortex
field is even, leading to abelian statistics only.
If formulated on an orientable manifold, the low energy theory can be understood as a $\Z_{2r}$ gauge theory
with Chern-Simons term, by arguments explained  in \cite{SW}.

For $\nu$ any multiple of 4, similar models are possible.  
The first case in which we can take all of the $\chi$ fields to have odd charge is $\nu=8$, which requires four $\chi$ fields.
For example, for $\nu=8$,
we can introduce four $\chi$ fields all of charge 1, and take $\phi$ to have charge 2.   The low energy theory is a $\Z_2 $ gauge theory
(with no Chern-Simons term, as it turns out).

\subsubsection{$\nu$ Divisible By 16}\label{sixteen}

$\nu=16$ is the first case in which the emergent gauge symmetry can be completely broken, as we learned in section \ref{overview}.
Accordingly, it is the first case that it might be possible to obtain a gapped phase that is also topologically trivial.  This is consistent with the fact that the
anomaly of the bulk topological superconductor is trivial at $\nu=16$.  Several constructions of $\sT$-invariant, gapped, topologically
trivial boundary states are known at $\nu=16$.

In the present approach, at $\nu=16$, we can take four Dirac fermion fields $\chi_i$ of charge 0 and four such fields  $\chi'_i$ of charge 1.
We also take $\phi$ to have charge 1.  The expectation value of $\phi$ completely breaks the gauge symmetry and a Yukawa
coupling $\sum_{i=1}^4\chi_i\chi'_i\bar\phi$ can give mass to all fermions.   A monopole operator that can be added to the action or
Hamiltonian completely destroys the conservation of vorticity.  The resulting low energy theory is gapped, symmetry-preserving, and topologically
trivial.

\subsubsection{Models At $\nu=16$ That Are Gapped Via Nonperturbative Effects}\label{confinement}

All of the models described so far can be understood entirely in the framework of weak coupling. The only effect beyond conventional perturbation theory was to add to the action
a monopole operator, with a small coefficient.  The only role of this operator was to explicitly break a symmetry, so it can itself be treated perturbatively.

It is interesting to also consider models that become gapped with the aid of nonperturbative effects.

For a first example, we take the gauge group to be $G=\U(2)$. The center of $G$ is thus a copy of $\U(1)$, embedded as $\mathrm{diag}(e^{\i\alpha},e^{\i\alpha})$.
 We introduce four doublets of Dirac fermions $\chi_1,\dots,\chi_4$, each transforming in the $\bf 2$ of $\U(2)$.  
The $\2$ is a two-dimensional complex representation or equivalently a four-dimensional real representation.   We have shown in section \ref{examples} that a model with fermions comprising four copies of the $\2$  has no gauge anomaly.
Since it has $\nu=4\times 4=16$, it also has no gravitational anomaly and is completely anomaly-free.

To make possible complete Higgsing of $\U(2)$ to the standard boundary state of the topological superconductor, we introduce $w$-fields consisting of two or more ($\sT$-even) doublets
of $\U(2)$.  To make possible a flow to a symmetry-preserving, gapped, topologically trivial state, we add another ($\sT$-odd) Higgs field $\phi$ that is invariant under
the subgroup $\SU(2)\subset G$ and transforms with charge 2 under the center of $G$.  This has been chosen to make possible gauge-invariant and $\sT$-conserving
Yukawa couplings:
\be\label{turner}\bar\phi\epsilon_{AB}\epsilon_{ab}\left(\chi_1^{Aa}\chi_2^{Bb}+\chi_3^{Aa}\chi_4^{Bb}\right)+\mathrm{h.c.}\ee
(Here $A,B=1,2$ and $a,b=1,2$ are respectively $\U(2)$ indices and spinor indices.)
Once $\phi$ gets an expectation value, the fermions become massive and the gauge symmetry is spontaneously broken from $\U(2)$ to $\SU(2)$.
In perturbation theory, therefore, one sees at low energy a pure $\SU(2)$ gauge theory, with no massless charged fields.  
In the infrared, though there is no explicit proof (except from numerical simulations), 
it is strongly believed that this theory flows to a confining, gapped, $\sT$-conserving,
and topologically trivial phase.  If so this gives another way to gap a $\nu=16$ system in a $\sT$-conserving way.

The reason that in this construction we had to start with an emergent $\U(2)$ gauge symmetry rather than just the center $\U(1)$ of $\U(2)$ is the following.
Had we gauged only the central $\U(1)$, then after the charge 2 Higgs field $\phi$ 
acquired an expectation value, $\U(1)$ would have been spontaneously broken to $\Z_2$.
The low energy physics would have been a $\Z_2$ gauge theory, 
rather like other examples that were discussed earlier.   The low energy time-reversal symmetry
would satisfy the familiar relation
\be\label{omigo}\slT^2=(-1)^\sF\sK,\ee
where $\sK$ is the nontrivial element of $\Z_2$.

How might we get rid of the $\Z_2$ gauge symmetry?  One idea would be to 
let a scalar carrying $\Z_2$ charge get an expectation value.  However, this would necessarily
break time-reversal symmetry, because eqn. (\ref{omigo}) shows that a scalar field that is odd under $\sK$ cannot be $\slT$-invariant.

An alternative might be to confine the $\Z_2$ gauge symmetry rather than Higgsing it.\footnote{The necessary dynamics is 
sometimes discussed under the name of ``vortex condensation.''
Note that in the present context, there is no local operator that creates a vortex of unit vorticity (in which the phase of 
$\phi$ changes by $2\pi$ and the flux of the broken
$\U(1)$ gauge field is $\pi$).  This would be a half-monopole operator, which does not exist.
The phrase ``vortex condensation'' can be understood as a reference to confinement of the gauge symmetry.}  
The trouble  is that although there is no problem to see confinement of a discrete
gauge symmetry such as $\Z_2$ in a lattice theory, it is hard to interpret this idea in a continuum field theory. 
One could take the emergent gauge symmetry $G$ to be defined on a lattice, but then how would we couple it to the fermions?
 Unfortunately, it is very difficult to explicitly construct
a $\sT$-invariant lattice fermion theory with nonzero $\nu$.  Because of the anomaly, one expects 
that it is only possible to do this (if at all)  if $\nu$ is a multiple of 16, and any
way to do it would probably itself lead immediately to a way to see that the $\nu=16$ model is 
topologically trivial.  Because of issues such as this, one approach
to gapping the $\nu=16$ problem in a $\sT$-invariant way has actually involved 
coupling lattice bosons to continuum fermions \cite{KitTwo}.  It is conceivable that this
could be adapted to the present context.

Instead of that, in the above construction,
we took the emergent gauge symmetry $G$ to be $\U(2)$ rather than its center $\U(1)$.  The effect of this was that the expectation value of $\phi$ broke
$G$ not to the finite group $\Z_2$ but to the connected Lie group $\SU(2)$, of which the $\Z_2$ in question is the center.  
Now to gap the system, we want confinement in pure $\SU(2)$ gauge theory and this
involves no problem, except that it is hard to understand.

There is a variant of this in which the emergent gauge group $G$ is taken to be not $\U(2)$ but $\U(1)\times \U(1)$, the maximal torus of $\U(2)$.  The fermion representation
is taken to be the same, viewed as a representation of the smaller gauge group, and the same Higgs fields $w$ and $\phi$ are introduced, with the same couplings.  The expectation
value of $\phi$ now breaks $G$ not to $\SU(2)$ but to the maximal torus of $\SU(2)$, which is the subgroup $\U(1)'=\mathrm{diag}(e^{\i\alpha},e^{-\i\alpha})$ of $\U(2)$.  The
low energy theory is a pure $\U(1)'$ gauge theory.  If we add to the Hamiltonian a monopole operator of unit magnetic charge 
with a small coefficient, then in the infrared we get the Polyakov model
\cite{Polyakov} of compact QED.  The infrared theory becomes confining, gapped, and topologically trivial.

This claim depends on  assuming that the monopole operator has magnetic flux 1 with respect to $\U(1)'$.  Otherwise, after confinement, we will be left over
with a nontrivial ``magnetic'' gauge group. The monopole operator must also satisfy Dirac quantization with respect to the
full underlying gauge symmetry $G$.  To satisfy these conditions, we simply take a monopole operator of charges $(1,0)$ (or $(0,1)$) in the original $\U(1)\times \U(1)$ gauge theory.

In this approach, the $\Z_2$ that potentially is left unbroken by the expectation value of a charge 2 Higgs field $\phi$ has been embedded in the connected group
$\U(1)'$ and then has been confined.

\subsection{Bosonic Analog}\label{bosonic}

\subsubsection{Goal}

An important part of the above story is the extension of symmetry groups
\be\label{extsym}1\to \K\to \H \to \T\to 1\ee
that is implied by the relation $\sT^2=(-1)^\sF\sK$.
Here as explained in section \ref{overview}, $\T$ is the bulk symmetry group generated by $\sT$, $\K$ is the group of emergent gauge symmetries on the boundary,
and $\H$ is a nontrivial extension of $\T$ by $\K$.
This group extension does not explain everything about these models, since for example it does not account for the nonabelian statistics that occur when $\nu$ is congruent
to 2 mod 4.  However, it accounts for a great deal, especially when $\nu$ is a multiple of 4.

It is therefore of interest that this part of the story has an
analog for purely bosonic symmetry-protected topological (SPT) phases.  Here we consider a 
purely bosonic theory in $D$ spacetime dimensions with a global symmetry group $\F$.  To keep
things simple, we will assume that $\F$ is a finite group of purely internal symmetries, not including reflection or time-reversal symmetries. 
We will describe boundary states for such theories constructed using group extensions analogous to (\ref{extsym}).

\subsubsection{Review}\label{realreview}

A unified description of a large class of SPT phases was presented in \cite{WenETAL} in terms of group cohomology.  These are theories with a global symmetry $\F$
that can be gauged\footnote{In the framework of \cite{WenETAL}, $\F$ is assumed to be an ``on-site'' symmetry, which ensures that it can be gauged.
There is a somewhat similar reason that $\sT$ in a topological superconductor can be gauged, in the sense that the effective description makes sense on an
unorientable spacetime.
The Schrodinger equation of electrons and nuclei makes sense on an unorientable spatial manifold, so
a $\sT$- and $\sR$-invariant phase of matter derived from it that has an
emergent relativistic symmetry that puts space and time on the same footing 
will always make sense on an unorientable spacetime.} (as opposed to theories with a global symmetry $\F$ that cannot be gauged because of an anomaly).
Gapped phases with a gaugeable $\F$ symmetry are then classified by their response to being gauged.  What this means is that one specifies the partition function 
the theory has if formulated on a compact, oriented $D$-manifold $X$, endowed with an $\F$-bundle (which is automatically flat, since $\F$ is assumed to be a finite group).

To specify the partition function in the presence of a background gauge field, the starting point 
is a cohomology class  in $H^D(\F,\U(1))$ that we will write as $e^{\i\upgamma}$.  
  For any  $X$ as above endowed with an $\F$-bundle, $e^{\i\upgamma}$ can be ``pulled back'' to a 
class $e^{\i\beta}\in H^D(X,\U(1))$.  (A slightly abstract explanation of this pullback operation is presented in the paragraph preceding
 eqn. (\ref{fibr}).) 
The class $e^{\i\beta}$ can be ``integrated'' over $X$, and this integral is supposed
to be the partition function.\footnote{Gauge theories of a finite gauge group in which the contribution of a particular flat bundle to the partition function
is the integral of $e^{\i\beta}$ (divided by the order of the automorphism group of the bundle) were constructed in \cite{DW}, without envisaging the application to condensed matter
physics.}  We write the integral evocatively as $Z_\beta= \exp(\i\int_X\beta)$.  
The motivation for the notation $e^{\i\int_X\beta}$ is that although what we are formally writing as $\int_X\beta$ is not the integral over $X$ of a gauge-invariant local effective action,\footnote{To be more
exact, there is no general, natural procedure to write it that way.  In a specific case, one may be able to add additional fields and gauge invariances
and write $\beta$ as some sort of local integral.} $Z_\beta$ has the
same formal properties as if it were.   For example, after making some gauge choices or introducing a more abstract language to avoid having to do this,\footnote{The more abstract language involves introducing a one-dimensional vector space $H_W$ of ``physical states'' for every boundary component
$W$ of $X$ ($H_W$ is one-dimensional because we are describing an SPT phase with no intrinsic topological order), 
and interpreting $\exp\left(\i\int_Y\beta\right)$ as a quantum mechanical transition amplitude between initial and
final states rather than a complex number.  Thus this more abstract language is actually essential in developing the theory more fully.
Nevertheless, we will not develop that language here.}
one can extend the definition of $e^{\i\int_X\beta}$ to the case that the manifold $X$ has a boundary, and then it satisfies a gluing law precisely of the form of the usual
gluing relation for the integrand of the Feynman path integral (eqn. (\ref{luth})), which as we have discussed also has a counterpart for the eta-invariant
(eqn. (\ref{gluth})).
  The gluing formula says that if an oriented $D$-manifold $Y^*$ is built by gluing
together two such manifolds $Y$ and $Y'$ along a component $W$ of their common boundary, as in fig. \ref{second} of section \ref{condensed}, then
\be\label{twerf}\exp\left(\i\int_Y\beta\right)\exp\left(\i\int_{Y'}\beta\right)=\exp\left(\i\int_{Y^*}\beta\right). \ee
The gluing formula is one way to understand the fact that $Z_\beta=\exp(\i\int_X\beta)$ is the partition function of a topological field theory.

  Likewise\footnote{Cobordism invariance of $\exp\left(\i\int_X\beta\right)$ means that if the oriented manifold $X$, equipped with an $\F$-bundle, is
  the boundary of an oriented manifold $Z$, over which the given $\F$-bundle extends, then $\exp\left(\i\int_X\beta\right)=1$.  This amounts to Stokes's theorem
  together with  $\d\beta=0$. (To be more exact, it follows from facts valid in any cohomology theory that reduce for differential forms
  to Stokes's theorem and $\d\beta=0$.)} $\exp\left(\i\int_X\beta\right)$ is a cobordism invariant (of an oriented $D$-manifold with an $\F$-bundle) rather as in even dimensions $\exp(-\pi\i\eta_R/2)$
is a cobordism invariant (of a $\pin^+$ $D$-manifold with a gauge bundle).   This cobordism property of $\exp(-\pi\i\eta_R/2)$ has not been made explicit in the present
paper; it can be deduced from the APS index theorem
\cite{APS},  as is explained in \cite{Witten}, section 4.2.
In general, a $\U(1)$-valued cobordism invariant is the partition function of an invertible topological field theory (see \cite{FM}, section 5.5, and \cite{FH}), and in particular cobordism
invariance implies 
that $\exp\left({\i\int_X\beta}\right)$ and $\exp(-\pi\i\eta_R/2)$ are the partition functions of suitable topological field theories.   That $\exp(-\pi\i\eta_R/2)$ is
the analog of  $\exp\left({\i\int_X\beta}\right)$ for a $\sT$-invariant   theory with fermions  satisfying $\sT^2=(-1)^\sF$ was the main idea\footnote{As explained
there, in some important examples, a more elementary description is possible in which one uses an ordinary index or a mod 2 index of the Dirac operator instead of the eta-invariant. The eta-invariant provides a general framework to understand all possible examples.} in \cite{Witten} (this was inspired by an earlier remark \cite{Ketal} concerning cobordism invariance of the eta-invariant in four dimensions).

\subsubsection{Gapped Boundaries For Bosonic SPT States}\label{gapbb}

To pursue the analogy, we will describe a simple class of gapped, symmetry-preserving boundary states for bosonic SPT phases that are analogous to the fermionic  states described in section \ref{uone}.  We consider a worldvolume $X$
of $D$ spacetime dimensions with a boundary $W$ of one dimension less. We assume that we are given a nontrivial group extension
\be\label{gext}1\to\K\to \H\to \F\to 1,\ee
where as before $\F$ is a (gaugeable) finite group of global symmetries of a system in $D$ spacetime dimensions, and $\K$ is a finite group of emergent gauge symmetries
that appear on $W$.   The reason that we assume that the symmetries in $\K$ are {\it gauge} symmetries is the following.  The symmetries in $\K$ have to be exact
symmetries or the construction that follows will not make sense.  In condensed matter physics, it is natural for {\it approximate} global symmetries to emerge
in the infrared; they are explicitly violated by interactions that are irrelevant in the renormalization group sense.   For gauge symmetries, the situation is the opposite;
it does not make much sense to have an emergent gauge symmetry in the infrared that is explicitly broken by irrelevant interactions.  On the contrary, exact emergent gauge
symmetries are often postulated in models of condensed matter systems, for example in models of the fractional quantum Hall effect.

How can we use the group extension (\ref{gext}) to construct a gapped boundary state?    Given a homomorphism $\H\to\F$, the class $e^{\i\upgamma}\in H^D(\F,\U(1))$
can be pulled back to a class $e^{\i\hat{\upgamma}}\in H^D(\H,\U(1))$. (See the discussion of eqn. (\ref{fibr}) for a slightly abstract explanation.) If $e^{\i\hat\upgamma}$ is trivial, then any trivialization of this class can be used to construct
a gapped boundary state, as we will explain.   Note, however, that if $e^{\i\upgamma}$ is nontrivial (so that the bulk SPT phase for which we are trying to
construct a gapped boundary state is nontrivial),  then 
to make  $e^{\i\hat\upgamma}$ trivial, the group extension (\ref{gext}) will have to be nontrivial.    (If the full
symmetry group is just a product $\K\times\F$, the existence of $\K$ will not help us trivialize a cohomology class of $\F$.)
 So to construct a gapped, symmetry-preserving boundary state in this way requires a nontrivial group extension at the starting point. 
For a simple concrete example, take $\F=\K=\Z_2$, $\H=\Z_4$.    These fit in a nontrivial extension:
\be\label{zooff}0\to\Z_2\overset{2}{\rightarrow} \Z_4\overset{r}\rightarrow\Z_2\to 0. \ee  The first map is multiplication by 2 and the second is reduction mod
2.  (Because the groups involved are all abelian, we have used an additive notation rather than the multiplicative notation in eqn. (\ref{gext}).)
This particular example is interesting in $D=3$ spacetime dimensions, since $H^3(\F,\U(1))\cong \Z_2$, and the nontrivial class 
$e^{\i\upgamma}$ in $H^3(\F,\U(1))$ is trivial when pulled back to
$e^{\i\hat\upgamma}\in H^3(\H,\U(1))$.  (In this example, $H^2(\K,\U(1))=0$, so the trivialization of $e^{\i\hat\upgamma}$ is essentially unique, and accordingly the following construction will give only one gapped boundary state.)

If $e^{\i\hat\upgamma}$ is trivial, then its trivializations form topological classes that differ by  elements of $H^{D-1}(\H,\U(1))$.
As we will see, any class of trivialization will give a gapped symmetry preserving boundary state for the topological field theory with partition function $e^{\i\beta}$.  
So the number
of such boundary states that can be constructed in this way (once the group extension (\ref{gext}) is chosen) is simply the order of the finite
group $H^{D-1}(\H,\U(1))$.
The physical meaning of this is straightforward. An element of  $H^{D-1}(\H,\U(1))$ determines a  purely $D-1$-dimensional invertible topological
field theory $\T$ with symmetry group $\H$.  Any boundary state we define can be modified by tensoring it with $\T$.   So if we can make
any boundary states at all, the number of such boundary states that we can make will be the order of $H^{D-1}(\H,\U(1))$.
What we have just said can be compared with what was explained in section \ref{interpretation}.  Triviality of $\exp(\i\hat\upgamma)$ is the analog of triviality of $\exp(-\pi\i
\eta_R/2)$.  And $H^{D-1}(\H,\U(1))$ is the analog of the finite group $\mathrm{Hom}(\Gamma',\U(1))$ that appeared in section \ref{interpretation}. 
Given that $\exp(\i\hat\upgamma)$ or $\exp(-\pi\i\eta_R/2)$ is trivial, the number of symmetry-preserving
gapped boundary states that can be constructed using this fact is the order of 
$H^{D-1}(\H,\U(1))$ or of $\mathrm{Hom}(\Gamma',\U(1))$.

We will now explain a few mathematical points that have been omitted so far. 
One way to describe group cohomology is that it is cohomology of the classifying space.  The classifying space $B\F$ of a group $\F$ is defined as the quotient
$E\F/\F$, where $E\F$ is a contractible space with a free action of $\F$.  (Such a space always exists, and any two choices are homotopic so
it does not matter which one is chosen.)
Then by definition $H^D(\F,\U(1))=H^D(B\F,\U(1))$.  The space $B\F$ is endowed with a principal $\F$-bundle, which is simply the total space of the fibration
$E\F\to B\F$.  This $\F$-bundle is ``universal,'' in the sense that if $X$ is any topological space with an $\F$-bundle, then that $\F$-bundle is the pullback to $X$
of the universal
$\F$-bundle over $B\F$ by some map $\Phi:X\to B\F$ (defined up to homotopy, and called the classifying map).  Accordingly, if we are given a class 
$e^{\i\upgamma}\in H^D(\F,\U(1))=H^D(B\F,\U(1))$, we can pull
this class back to $e^{\i\beta}\in H^D(X,\U(1))$.  This is the pullback operation that we used in section \ref{realreview} in the initial
construction of an SPT phase based on the cohomology class $e^{\i\upgamma}$.
On an oriented  manifold $X$ without boundary, the partition function of the SPT phase associated to $e^{\i\upgamma}\in H^D(\F,\U(1))$
is then $\exp(\i\int_X\beta)$.

Now let us consider the group extension (\ref{gext}).  By definition, $\H$ acts freely on the contractible space $E\H$.  Since the group extension $\H$
entails a homomorphism $\H\to \F$,  $\H$ also acts on any space that $\F$ acts on, and in particular $\H$ acts (but not freely) on $E\F$.
Consequently, $\H$ acts on $E\H\times E\F$, and this action is free because it is free on the first factor.  Hence we can define
$B\H=(E\H\times E\F)/\H$.  By forgetting the first sector, $B\H$ projects to $E\F/\H=E\F/\F=B\F$.   The fiber is $E\H/\K$, where $\K$ (like all of $\H$)
acts freely on $E\H$, so we can set $E\H/\K=B\K$.  Thus there is a fibration
\be\label{fibr}\begin{matrix} B\K&\overset{i}{{{\longrightarrow}}} &B\H\cr
  & &\Big\downarrow \pi \cr
 &  & B\F.\end{matrix}\ee
 
Now we can describe the ``pullback'' operation that is part of the definition of the boundary state.   Given $e^{\i\upgamma}\in H^D(B\F,\U(1))$, we simply
 pull it back to $e^{\i\hat\upgamma}=\pi^*(e^{\i\upgamma})\in H^D(B\H,\U(1))$.  
 
 We are almost ready to define the boundary state.
To describe a boundary state for a theory whose partition function would be $e^{\i\int_X\beta}$ if $X$ has no boundary,
we  have to describe what is meant by $e^{\i\int_X\beta}$ when $X$ has a boundary. 
There is no symmetry-preserving definition without introducing some new variables on the boundary.
  On what sort of boundary data will $e^{\i\int_X\beta}$ depend?
The bulk physics on $X$ has $\F$ symmetry, and we assume it has been coupled to some background $\F$-bundle (possibly trivial). This means
that the bulk theory on $X$ has been coupled to a background gauge field of the finite group $\F$.
However, on $W$  we postulate the existence of fields 
 whose role is to ``lift'' the background $\F$-bundle (or more precisely its restriction to $W$) to an $\H$-bundle.
(These fields are analogous to $\phi$ in the fermionic models that were introduced in section \ref{overview} above.)   Accordingly,
we assume that on $W$ we are
given an $\H$-bundle which, 
  under the projection $\H\to \F$ that is part of the extension (\ref{gext}), is mapped to (the restriction to $W$ of) the background $\F$-bundle.
  The definition of $\exp(\i\int_X\beta)$ will depend on this $\H$-bundle.
  
A given $\F$-bundle over $X$, when restricted to $W$, may not lift to $\H$ at all.  Such a bundle contributes 0 to the path integral for the boundary state that we will define; it is not compatible with this boundary state.  Alternatively, a given $\F$-bundle, restricted to $W$, may lift to $\H$ in more than
one way.  If $\H$ is abelian, then any liftable $\F$-bundle can be lifted in precisely $N$ ways, where $N$ is the order of the finite group $H^1(W,\K)$.
There is no equally simple
statement if $\H$ is not abelian.   Regardless, assuming that the background $\F$-bundle over $X$ is liftable to $\H$, we compute the partition function, in the presence
of the boundary state on $W$, by summing over the possible lifts, with the contribution of each lift being $\frac{1}{n}e^{\i\int_X\beta}$; here $n$ is the order\footnote{The reason
to divide by $n$ is that $\K$ is a group of gauge symmetries.  In quantizing a gauge symmetry, one always has to divide by the volume of the automorphism group.}
of the subgroup of $\K$ that acts by automorphisms on a given lift, and we explain next what is meant by $ e^{\i\int_X\beta}$.   That will complete the definition of
the boundary state.

To integrate the cohomology class $e^{\i\beta}$ over a manifold $X$ with boundary, we need a trivialization of this class over $W=\partial X$.
That is precisely what we have if we are given a trivialization of $e^{\i\hat\upgamma}$.  Indeed, given the background $\F$-bundle over $X$ and a lift to $\H$
of the restriction of
this bundle to $W$, we have the following commutative diagram:
\be\label{diagramm}\begin{matrix}  W & \overset{j}{\longrightarrow }& X \cr
                                                         \Big\downarrow\hat\Phi&& \Big\downarrow\Phi\cr
                                                          B\H & \overset{\pi}{\longrightarrow} & B\F.       \end{matrix}\ee
 Here $j$ is the inclusion of $W$ in $X$, $\pi$ was described in eqn. (\ref{fibr}), the background $\F$-bundle over $X$ is the pullback by  $\Phi$ of the universal
 $\F$-bundle over $B\F$, and the lifted $\H$-bundle over $W$ is similarly a pullback by $\hat\Phi$ of the universal $\H$-bundle over $B\H$.
  Since by definition $e^{\i\beta}=\Phi^*(e^{\i\upgamma})$,
 the restriction to $W$ of $e^{\i\beta}$ is $j^*\Phi^*(e^{\i\upgamma})$.  Commutativity of the diagram means that this is the same as $\hat\Phi^*\pi^*(e^{\i\upgamma})
 =\hat\Phi^*(e^{\i\hat\upgamma})$.  Hence any trivialization of $e^{\i\hat\upgamma}$ gives a trivialization of     the restriction to $W$ of $e^{\i\beta}$, and enables
 us to define $e^{\i\int_X\beta}$.

 \subsubsection{The Boundary Topological Field Theory}  
 
This construction has at least one more important facet.  Going back to the fibration (\ref{fibr}), we see that $B\K$ is the fiber of the fibration $B\H\to B\F$, and thus
 we can think of $B\K$ as the inverse image in $B\H$ of a point  $p_0\in B\F$.   The class $e^{\i\upgamma}\in H^D(B\F,\U(1))$ is  trivial when restricted to a point,
 and this gives it a natural trivialization when pulled back to $B\K$.  In other words, the restriction to $B\K$ of $e^{\i\hat\upgamma}\in H^D(B\H,\U(1))$
 is naturally trivial, whether $e^{\i\hat\upgamma}$ is trivial or not.  On the other hand, if we are given a global trivialization of $e^{\i\hat\upgamma}$, we can restrict
 this trivialization to $B\K$, and therefore we now have two trivializations of the restriction to $B\K$ of $e^{\i\hat\upgamma}$.  The ratio of two trivializations of
 $e^{\i\hat\upgamma}|_{B\K}\in H^D(B\K,\U(1))$ is a class $e^{\i\updelta}\in H^{D-1}(B\K,\U(1))$.  
 
Such a class is precisely what we need to describe a version of $\K$ gauge theory on a $D-1$ dimensional spacetime
 manifold such as $W$.  This theory is a nontrivial topological field theory, and we claim that
 it is realized along $W$ in the boundary state that we have described.  The justification for this
 statement is as follows.  Let us take the background $\F$-bundle over $X$ to be trivial.  Then its restriction to $W$ is, of course, also trivial.  For a lift of the trivial
 $\F$-bundle to $\H$, we can simply take any $\K$-bundle over $W$. In this situation, we can take  $\Phi:X\to B\F$ to map $X$ to a point, which we can take to
 be $p_0$.  
 Then $e^{\i\beta}=\Phi^*(e^{\i\upgamma})$ is naturally trivial along $X$, and in particular along $W$, since a pullback of a class of positive degree by a constant
 map is always naturally trivial. This
trivialization agrees along $W$ with the one that comes from the fact that $e^{\i\hat\upgamma}$ is naturally trivial when restricted to $B\K$.  
(They are both derived from the fact that $e^{\i\upgamma}$ is trivial when restricted to a point.)
We can further take $\hat\Phi:W\to B\H$ to be a map to $B\K\subset B\H$. 
If we  use the trivialization of $e^{\i\beta}$ along $W$ that comes from constancy of $\Phi$ to define $ e^{\i\int_X\beta}$, we get the answer 1, because this trivialization extends
over all of $X$.   We are instead supposed
 to use along $W$ the other trivialization of $e^{\i\beta}$ -- the one that  comes by restricting to $B\K$  the global trivialization of $e^{\i\hat\upgamma}$ that was used
 to define the boundary state.
 By definition, the two differ by $e^{\i\updelta}$ and therefore $e^{\i\int_X\beta}$ computed using the second trivialization of $e^{\i\beta}$ along $W$
 is equal to $e^{\i\int_W\updelta}$.

So our boundary state can be defined by taking on $W$ a $\K$ gauge theory defined with the cohomology class $e^{\i\updelta}$ and
coupling it  the global symmetry $\F$ in bulk.     This coupling uses the class $e^{\i\upgamma}\in H^D(B\F,\U(1))$ and a trivialization of the pullback of this class
 to $B\H$.  This coupling of the $D-1$-dimensional boundary gauge theory to the global symmetry $\F$ is anomalous in the sense
that it cannot be defined in $D-1$ dimensions.  A purely $D-1$-dimensional coupling of the boundary gauge theory to an $\F$ global symmetry would
be made using a class $e^{\i\upgamma'}\in H^{D-1}(B\H,\U(1))$ whose restriction to $B\K$ is $e^{\i\updelta}$.  This gives what
 is known as a symmetry-enriched topological (SET) phase of matter \cite{Wen,EH,MR,HW}.\footnote{The idea behind the name is that a nontrivial topological field theory in $D-1$ dimensions
-- the $\K$ gauge theory defined with the cohomology class $e^{\i\updelta}$ -- has been enriched with a global symmetry $\F$.  Note that since
$\K$ is a group of gauge symmetries, the global symmetry group that acts on gauge-invariant excitations in a model of this type is indeed precisely $\F$.  An SET phase
reduces to an SPT phase if the topological field theory is trivial.}
Obviously, the type of theory we have described is not a purely $D-1$-dimensional theory of this kind.  It was obtained by a sort of anomalous coupling
of a topological field theory on the boundary of a $D$-manifold to an SPT phase in the bulk of that $D$-manifold,   and the starting point was a class in
$H^D(B\F,\U(1))$, not in $H^{D-1}(B\H,\U(1)).$   (It has been suggested by J. Wang  that the model in \cite{CBVF} may be an example of the class of states described here.)

We started this analysis with the bulk physics -- controlled by a class  $e^{\i\upgamma}\in H^D(B\F,\U(1))$ -- and not the boundary topological field theory
-- controlled by a class $e^{\i\updelta}\in H^{D-1}(B\K,\U(1))$ -- because this made the analysis much more straightforward.

\section{Application To M2-Branes}\label{mtheory}

M-theory is defined on an eleven-dimensional spacetime manifold that we will call $Y$.  $Y$ is not necessarily orientable; in general,\footnote{Stieffel-Whitney classes $w_k(V)$ (which entered our analysis in sections \ref{addf} and \ref{examples}) are defined for any real vector
bundle $V$ over a space $Z$.
If $Z$ is a manifold and $TZ$ is its tangent bundle, one writes $w_k(Z)$ (or just $w_k$ if the context is clear) for $w_k(TZ)$.}
  $w_1(Y)\not=0$.
However, $Y$ is a $\pin^+$ manifold and in particular
\be\label{zoff}w_2(Y)=0.\ee

M-theory also has ``membranes,'' known as M2-branes.  An M2-brane is a two-dimensional 
object; at a given time, it fills out a two-manifold $\Sigma$ in space. ($\Sigma$ has no boundary
unless $Y$ itself has a boundary or  one considers M5-branes; we will not consider these possibilities here.)  Taking the time into account, the
 worldvolume of an M2-brane is a three-manifold that we will call $W$.  As usual, in analyzing anomalies, it is convenient to work in Euclidean
signature and to take $W$ to be compact.   We write $TW$ for
the tangent bundle to $W$ and $N$ for its normal bundle in $Y$.  $N$ has rank $11-3=8$. 

\subsection{The Classical Picture}\label{classpict}

We will first describe M2-branes at the classical level, and then explain how this picture is modified by a quantum anomaly.

There is no restriction at all on the topology of $W$, which can be a completely general three-manifold.  However, there is a topological
restriction on how $W$ is embedded in $Y$:
its normal bundle  $N$ must be  oriented. 
Orientability of $N$ is equivalent to
\be\label{delmo} w_1(N)=0. \ee
We stress, however, that $N$ is not just orientable but oriented: it comes with a choice of orientation.
The orientation of $N$ determines whether
what is wrapped on $W$ is an M2-brane or an M2-antibrane.\footnote{If the $C$-field, which we introduce momentarily, were an
ordinary rather than twisted three-form field, then $W$ would have to be orientable and the difference between wrapping an M2-brane or
an M2-antibrane on $W$ would involve a choice of orientation of $W$, which
would determine the sign of $\int_WC$.   Because $C$ is a twisted three-form, it is the normal bundle to $W$ that has to be orientable,
and whose orientation determines whether what is wrapped is an M2-brane or M2-antibrane.}
(By convention, we will call the wrapped
object an M2-brane.)  

The tangent bundle of $Y$ decomposes along $W$ as $TY|_W= TW\oplus N$ (where the symbol $|_W$ represents restriction to $W$).
Together with the Whitney sum formula (\ref{sumf}) and $w_1(N)=w_2(Y)=0$, this implies that
\be\label{welmo}w_1(W)=w_1(Y)|_W,\ee
and 
\be\label{belmo}  w_2(N)=w_2(W). \ee
Eqn. (\ref{welmo}) implies that $W$ is orientable if and only if $Y$ is. 
 The orientation of $W$ is reversed in going
around a loop $\ell\subset W$ if and only if the orientation of $Y$ is reversed in going around that loop.
 But more specifically, since $N$ is oriented (and not just orientable),
 a local orientation of $Y$ determines a local orientation of $W$.
A fancy way to describe the situation is to say that the orientation bundle of $Y$ restricts along $W$ to the orientation bundle\footnote{\label{uprho} We can think of the orientation bundle as a principal $\Z_2$-bundle $\upzeta$ over $Y$ whose local sections correspond to local
choices of orientation of $Y$.  The holonomy of $\upzeta$
 around an orientation-reversing loop $\ell\subset Y$
is (in multiplicative notation) $-1$ while its holonomy around an orientation-preserving loop is $+1$.  Later, we introduce a twisted
sheaf of integers $\ti \Z$.  Once one picks a local orientation, $\ti\Z$ is equivalent to the constant sheaf $\Z$ of integers, but reversal of orientation
acts as $-1$ on $\ti\Z$.  A fancy definition is $\ti\Z=\upzeta\times_{\Z_2} \Z$ where $\Z_2$ acts on
$\Z$ as multiplication by $-1$.  If here one replaces $\Z$ with a trivial real line bundle $\o$, again with $\Z_2$ acting on $\o$ as multiplication
by $-1$, one can define a real line bundle $\varepsilon=\upzeta\times_{\Z_2}\o$.  It can be identified topologically with $\det TY$ and will play a role in section \ref{consistency}. } 
 of $W$.

M-theory also has a field $C$ that roughly speaking is a three-form gauge field (twisted in a sense that we will discuss).  We will first describe the properties of this field at the classical level and then describe
how they are modified by a quantum anomaly.

The $C$-field has a gauge invariance $C\to C+\d\Lambda$ where $\Lambda$ is a two-form, 
 and it has a gauge-invariant four-form field strength $G=\d C$.   $C$ is a three-form analog of an abelian gauge field $A$, which locally
is a one-form that has a gauge-invariance $A\to A+\d\phi$ and that has a gauge-invariant two-form field strength $F=\d A$.  In addition, ``big'' gauge transformations are allowed that
shift the periods of either $A$ or $C$ (i.e. their integrals over closed 1-manifolds or 3-manifolds) by integer multiples of $2\pi$.   (In the one-form case,
allowing ``big'' gauge transformations amounts to saying that the gauge group is $\U(1)$ rather than $\R$.  This has a three-form analog.)
  
A crucial difference is that $C$ is a ``twisted'' three-form, twisted by the orientation bundle
of $Y$.  This means that a choice of a local orientation of $Y$ turns $C$ into an ordinary three-form, but 
in going around an orientation-reversing loop $\ell\subset Y$, $C$ comes back with the opposite sign.  An ordinary three-form on $Y$ could be integrated
on an oriented three-dimensional submanifold $Q\subset Y$, but a three-form that is twisted in the sense just described can be integrated instead
on a submanifold $W$ with the property stated just after eqn. (\ref{belmo}) (a local orientation of $Y$ determines a local orientation of $W$, so that $C$ changes sign whenever
the orientation of $W$ is reversed).  
Thus we can consider an integral $\int_W C$, where $W$ is an M2-brane
worldvolume.  We call this (if $W$ is compact) a ``period'' of $C$.  But as the period of $C$ can be shifted by an integer multiple of $2\pi$ by a ``big'' gauge transformation,
the natural quantity is really the exponential $\exp(\i\int_W C)$.    This exponential is analogous to the holonomy $\exp(\i\oint_\ell A)$ of an abelian gauge field
$A$ around a closed loop $\ell$. 

  Because $C$ is twisted, its curvature $G=\d C$ is not an ordinary four-form but a twisted one, which changes sign in going around an orientation-reversing loop.

\begin{figure}
 \begin{center}
   \includegraphics[width=3.5in]{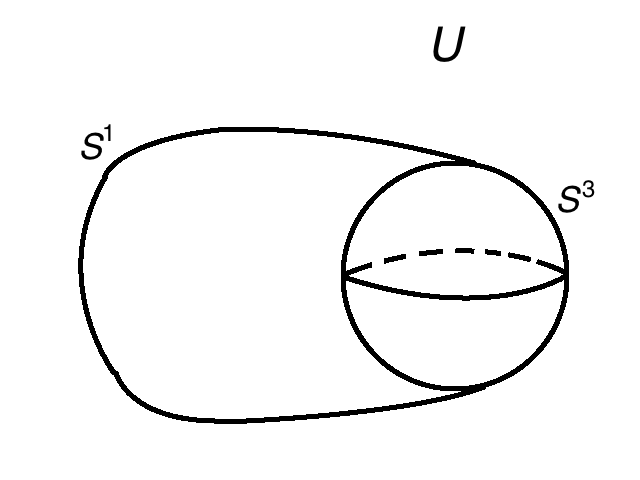}
 \end{center}
\caption{\small  Schematic depiction of a four-manifold $U=S^3\times S^1$.}
 \label{loop}
 \end{figure}

The M2-brane is ``charged'' with respect to $W$, meaning that it couples to $C$ via  a factor $\exp(\i\int_W C)$.  For this coupling to be well-defined
in the absence of fermion anomalies, $G$ must obey an analog of Dirac quantization of magnetic flux.  This states that if $U\subset Y$ is any
suitably twisted four-cycle,\footnote{That is, $U$ must have the same property as $W$: its normal bundle is oriented, so that the orientation bundle
of $Y$ pulls back along $U$ to the orientation bundle of $U$.}  then the flux on $U$ of  $G/2\pi$ must be an integer:
\be\label{derf}\int_U\frac{G}{2\pi}\in\Z. \ee
The reason for this condition is 
similar to the reason for Dirac quantization in abelian gauge theory.  Consider, for example, the case that $U=S^3 \times S^1$,
as is schematically shown in fig. \ref{loop}.  Let $W=S^3\times p$ with $p$ a point in $S^1$.  We want the coupling of $W$ to the $C$-field,
namely $\exp(\i\int_W C)$, to be single-valued when $p$ makes a loop around $S^1$ and returns to the starting point.  The condition for this is
\be\label{erf}\exp\left(\i\int_{S^3\times S^1}G\right)=1,\ee
which is equivalent to the flux quantization condition (\ref{derf}).

The analogy between the three-form field $C$ and an abelian gauge field $A$ can be further extended as follows.
An abelian gauge field on a manifold $Y$ (or more accurately the complex line bundle on which the gauge field is a connection)
has a characteristic class $c_1$ (the first Chern class), 
 valued in $H^2(Y,\Z)$.  At the level of differential forms, $c_1$ is represented by the de Rham cohomology class of $F/2\pi$,
but $c_1$ also contains torsion information that is not captured by de Rham cohomology.
The $C$-field has an analogous characteristic class that we will call $x$.  Modulo torsion, $x$ is represented by the de Rham cohomology class of $G/2\pi$,
which is integral by virtue of Dirac quantization.
However, as $G$ is a twisted four-form, $x$ takes values not in the ordinary cohomology $H^4(Y,\Z)$ but in a twisted cohomology group that we will call
$H^4(Y,\ti \Z)$.  Here (see footnote \ref{uprho})
$\ti\Z$ is a twisted sheaf of integers over $Y$, twisted by the orientation bundle of $Y$.  The assertion that $x$ takes values in this twisted
cohomology rather than in the ordinary cohomology of $Y$ just reflects the fact that $C$ and $G$ are not ordinary differential forms but twisted ones.  

The topological choice of an abelian gauge field on a manifold $Y$ is completely classified by the first Chern class, which can be an arbitrary element of $H^2(Y,\Z)$,
and likewise the topological choice of the M-theory $C$-field is completely classified, at the classical level, and ignoring the fermion anomaly,
 by the characteristic class $x$, which can be an
arbitrary element of $H^4(Y,\ti \Z)$.

\subsection{Worldvolume Fermions And The Anomaly}\label{wv}

The M2-brane volume $W$ also supports fermions.  Roughly speaking, these fermions
are spin 1/2 fermions on $W$ with values in positive chirality spinors of the normal bundle $N$ to $W$.

A more precise description is as follows.  Since the full spacetime manifold $Y$ is a $\pin^+$ manifold, it is endowed with a $\pin^+$ bundle
$P\to Y$.   In 11 dimensions, the rank of $P$ is 32.

Since the normal bundle $N$ to $W$ in $Y$ is oriented and of rank 8, it is endowed with an antisymmetric Levi-Civita tensor $\epsilon_{j_1j_2\dots j_8}$.
At a point $p\in Y$, there are Dirac gamma matrices $\G_I,\,\,I=1,\dots ,11$ acting on $P$.  For $p\in W$, we can split the $\Gamma_I$
into a triplet $\gamma_i$, $i=1\dots 3$ that are tangent to $W$, and an additional eight gamma matrices $\ti\g_j$, $j=1,\dots,8$ that are normal to $W$.
We introduce the chirality operator of $N$:
\be\label{zolbo}\h\Gamma=\frac{1}{8!}\epsilon^{j_1\dots j_8}\ti\g_{j_1}\ti\g_{j_2}\dots \ti\g_{j_8}. \ee

The M2-brane worldvolume fermion field $\Psi$ is a  section of $P$, or more exactly  of $P|_W$ (the restriction of $P$ to $W$) that obeys
\be\label{olbo}\h\Gamma \Psi=\Psi. \ee
This can be described roughly by saying that $\Psi$ is a spinor on $W$ with values in positive chirality spinors of $N$, but that
is only a rough description since in general $w_2(W)$ and $w_2(N)$ are both nonzero (they are equal as in eqn. (\ref{belmo})),
in which case there exist neither spinors of $W$ nor spinors of $N$.  However, given the $\pin^+$ structure of $Y$ and the orientation of $N$,
there is never a problem in defining the bundle $P_+\to W$ whose sections are sections of $P|_W$ that satisfy eqn. (\ref{olbo}).
The M2-brane fermions are sections of $P_+$.  Note that the rank of $P_+$ is $\frac{1}{2}\cdot 32=16$.  The Dirac operator that acts on the M2-brane fermions
is the obvious $\D=\i\slashed{D}=\i \gamma^i D_i$, where $D_i$ is the Riemannian connection on $P$, projected to $P_+$.

The reason that the classical picture of section \ref{classpict} requires modification is that the path integral of the fermion field $\Psi$ on $W$
has anomalies.  To orient the reader, we will describe these anomalies in two opposite cases.

First, let us assume that $w_2(W)=0$, but allow for the possibility that $w_1(W)\not=0$.  Since $w_2(W)=0$, we have $w_1(N)=w_2(N)=0$
(eqns. (\ref{delmo}) and (\ref{belmo})).  That means that $N$ is topologically trivial.\footnote{As in the discussion of eqn. (\ref{zork}), a real vector
bundle $N$ over a three-manifold $W$ is stably classified by $w_1(N)$ and $w_2(N)$.  If the rank of $N$ is large enough to be in the ``stable
range,'' then the stable classification of $N$ is the same as a topological classification.  For $W$ of any dimension,
a real vector bundle  $N$ is in the stable range if its rank exceeds the dimension of $W$.}  Let us then restrict to the special case that $Y=W\times \R^8$.
In this case, the $\pin^+$ structure on $Y$ determines a $\pin^+$ structure $P_W$ over $W$, and the field $\Psi$ just consists of eight Majorana
fermions all valued in $P_W$.  

This is a familiar system in the theory of topological superconductors.  Consider a topological insulator with the value $\nu=8$ of the usual
mod 16 invariant.  The standard boundary state of this system consists of 8 Majorana fermions, all taking values in the $\pin^+$ bundle of the boundary.
In general, for any value of $\nu$, a system of $\nu$ Majorana fermions on a possibly unorientable three-dimensional $\pin^+$ manifold $W$
has a partition function that is only well-defined up to a power of $\exp(-2\pi\i\nu/16)$ \cite{R,Witten}.  For $\nu=8$, this means that the partition
function is only well-defined up to sign.  Something must cancel this anomaly.  As will become clear, the anomaly cancellation mechanism is similar
in spirit to what happens in a topological superconductor, though different in detail.

For an example of an opposite kind, suppose that $Y$ and therefore also $W$ is orientable and pick orientations.    Then the $\pin^+$ structure of $Y$ becomes a spin
structure.  Suppose further that
$w_2(W)=w_2(N)=0$.  
One can then pick a spin structure on $W$, with spin bundle $\S$,
 and once this is done, the normal bundle $N$ to $W$ also acquires a spin structure.
 This means that one can define the spinor bundle of $N$, which we call $\S(N)$ ($\S(N)$ is the vector bundle over $W$
 that is associated to $N$ via the spinor representation of $\Spin(8)$). 
In this situation, the spin bundle of $Y$, restricted to $W$, is $\S\otimes \S(N)$.
 Moreover, $\S(N)$ has a decomposition $\S(N)=\S_+(N)\oplus \S_-(N)$
 in eigenspaces of the chirality operator $\h\Gamma$ (eqn. (\ref{zolbo})).  Eqn. (\ref{olbo}) now means  that $\Psi$ is
 valued in $\S\otimes \S_+(N)$, that is, it is a spinor on $W$ valued in positive chirality spinors of the normal bundle $N$.
 
From the point of view of fermion anomalies, $\Psi$ is simply a fermion field on $W$ coupled to a $\Spin(8)$ gauge field.  The $\Spin(8)$
gauge field is  the Levi-Civita connection of the normal bundle $N$ of $W$, lifted from $\SO(8)$ to $\Spin(8)$ via the choice of spin structure.  The microscopic
origin of the effective gauge field as part of the gravitational connection in higher dimensions does not affect the analysis of the fermion anomaly,
so we can borrow standard results.
The spinor representation of $\Spin(8)$ is subject to the standard ``parity'' anomaly, which is a problem in defining the sign of the fermion path integral.
An anomaly cancellation mechanism is again needed.

A concrete example of this type in which the ``parity'' anomaly affects the sign of the M2-brane path integral is as follows.  We return to the
example of  fig. \ref{loop}, with $Y$ containing an embedded copy of $U=S^3\times S^1$. 
Assume that $Y$ is the total space of a rank 7 
vector bundle $Q$ over $U$.
The structure group of $Q$ is $\SO(7)$, which lifts to $\Spin(7)$ via the choices of spin structure,
and we take it to have instanton number 1 (or more generally any odd instanton number). 
As before, we take $W$ to be $S^3 \times p$, with $p$ a point in $S^1$.
As $p$ is drawn in a loop around $S^1$ and returns to its starting point, the spectrum of the Dirac operator on $W$ undergoes a spectral flow and
the fermion path integral on $W$ changes sign.  This sign change is the usual ``parity'' anomaly, whose description via spectral flow can be found
 in section 2  of \cite{Witten}.

It is notable that in the two examples we considered, the anomaly only affects the sign of the M2-brane fermion path integral. 
In general, a fermion path integral in a spacetime of odd dimension has only discrete, ``global'' anomalies.  In dimension 3, the anomaly is in general
a $16^{th}$ root of 1.  But for the M2-brane, the anomaly in general is just a sign $\pm 1$, as we will see.

\subsection{Anomaly Cancellation On An Orientable Manifold}\label{zerob}

In \cite{Wittenold}, a natural anomaly cancellation mechanism was described for the case that $Y$ is orientable.  This mechanism was also shown
to solve several other problems (such as cancelling an anomaly in the overall M-theory path integral).

We will explain this mechanism in the context of the example of fig. \ref{loop}, and later generalize it.  Anomaly cancellation
depends on combining  the fermion
path integral, which we formally denote as $\Pf(\slashed{D})$ (where $\slashed{D}$ is the Dirac operator that acts on $\Psi$ and  $\Pf$ is the Pfaffian)
with the classical coupling of $W$ to the $C$-field.  Formally the product of the two is
\be\label{donu}\Pf(\slashed{D})\exp\left(\i\int_W C\right). \ee
It is this product that must be well-defined, rather than either factor separately.  
So if $\Pf(\slashed{D})$ changes sign when $p$ is taken around a loop in $S^1$, we want $\exp\left(\i\int_W C\right)$ to likewise change sign.
In eqn. (\ref{derf}), we imposed an integrality condition on $\int_{S^3\times S^1}G/2\pi$ in order to ensure that $\exp\left(\i\int_WC\right)$ is well-defined.  However, now we want $\exp\left(\i\int_WC\right)$ to change sign when $W$ goes around the loop precisely when the
instanton number on $U$, which we will call $\varrho$, is odd.  This will happen if $\int_U G/2\pi$ is a half-integer when the instanton number is odd:
\be\label{zases}\int_U\frac{G}{2\pi}=\frac{\varrho}{2}~~{\mathrm{mod}} ~\Z.\ee

This condition can be reformulated as follows.  Since  $Y$ is orientable in the example under discussion, the $\pin^+$ structure of $Y$ reduces
to a spin structure.  On a spin manifold $Y$, there is a four-dimensional characteristic class\footnote{$\lambda$ does not coincide with $p_1(Y)$,
the first Pontryagin class of $Y$.  Rather, the relation between them is $p_1=2\lambda$.  On a spin manifold, $p_1$ can be divided by
$2$ in a canonical way and we write $\lambda$ for $p_1/2$. Modulo torsion, $\lambda$ can be characterized by the condition on the instanton number
that is mentioned shortly.}   $\lambda$ that measures the instanton number of the tangent bundle, in the
sense that for any oriented four-manifold $U\subset Y$ without boundary, the instanton number of the tangent bundle of $Y$, integrated over $U$, is
$\int_U\lambda$.  Thus the condition (\ref{zases}) can be written
\be\label{zasess}\int_U\frac{G}{2\pi}=\frac{1}{2}\int_U \lambda~~{\mathrm{mod}}~\Z,\ee
and we impose on $G$ this modified quantization condition (for  any embedded four-manifold $U$ without boundary).
It is shown in eqn. (\ref{donu}) that for $Y$ orientable, the M2-brane path integral is well-defined if eqn. (\ref{zasess}) is satisfied.  Our main
goal in what follows is to generalize this analysis to the case that $Y$ is not necessarily orientable.  (We will also make some technical improvements
in the derivation even in the orientable case.)

The shifted quantization condition (\ref{zases}) has a one-form analog that is possibly more familiar.  If $A$ is a $\U(1)$ gauge field
with field strength $F=\d A$, then the periods of $F/2\pi$ are integers.  Suppose instead that $A$ is not an ordinary $\U(1)$ gauge field, but a $\spinc$
connection (a notion that was important in \cite{SW}).  Then, on a manifold $Z$ that is not necessarily a spin manifold, $F$ obeys a shifted
version of Dirac quantization.  For any two-cycle $V\subset Z$,
\be\label{yases}\int_V\frac{F}{2\pi}=\frac{1}{2}\int_V w_2(Z)~~{\mathrm{mod}}~\Z.\ee

To make the analogy even closer, we note that the mod 2 reduction of $\lambda$ is the Stieffel-Whitney class $w_4$.
Hence we can reformulate (\ref{zases}) as follows:
\be\label{ozases}\int_U\frac{G}{2\pi}=\frac{1}{2}\int_U w_4(Y)~~{\mathrm{mod}}~\Z.\ee
Thus the shifted quantization condition means that $C$ is a sort of three-form analog of a $\spinc$ connection.    The formulation (\ref{ozases}) has the advantage
of making sense on an unorientable manifold.  (The condition (\ref{zasess}) does not make much sense if $Y$ is unorientable, because in that
case $\lambda$ can be integrated on an embedded submanifold $U\subset Y$ if $U$ is oriented, while $G$ can be integrated on $U$ if $U$ has oriented normal bundle.)

The notion of a $\spinc$ connection can be stated somewhat more precisely as follows.  Suppose that $A$ is a $\spinc$ connection.  Then $2A$
is an ordinary $\U(1)$ gauge field, with field strength $2F$.  In terms of $2F$, the condition (\ref{yases}) reads
\be\label{yasess}\int_U\frac{2F}{2\pi}=\int_V w_2(X)~~{\mathrm{mod}}~2.\ee
This may be interpreted as follows.  The gauge field $2A$ has a first Chern class that we will call $\h c_1$, represented in de Rham cohomology by
$2F/2\pi$.  We interpret eqn.  (\ref{yasess}) to mean
that the mod 2 reduction of $\h c_1$ is $w_2$:
\be\label{wases} \h c_1\cong w_2~{\mathrm{mod}}~2. \ee This is often described by saying that  $\h c_1$ is an integer lift of $w_2$.
A $\spinc$ structure, for our purposes, is the topological class of a $\spinc$ connection.
Similarly, if $G=\d C$ obeys the modified quantization law (\ref{ozases}), then $2C$ is a conventional twisted three-form gauge field, 
with a characteristic class 
$\h x$, represented in de Rham cohomology by $2G/2\pi$.  We interpret eqn. (\ref{ozases}) to mean that the mod 2 reduction of $\h x$ is $w_4$:
\be\label{owases} \h x\cong w_4~{\mathrm{mod}}~2. \ee  So $\h x$ is a twisted integer lift of $w_4$ (a lift to the twisted integer cohomology 
$H^4(Y;\ti\Z)$).
By analogy with the  terminology  ``$\spinc$ connection,'' and for lack of any other name,
we will refer to an M-theory $C$-field that obeys the condition (\ref{owases}) and enables one to define the M2-brane path integral as an $\mc$ connection.
By an $\mc$ structure, we will mean the topological class of an $\mc$ connection.\footnote{What we have said so far is less than a full definition
of an $\mc$ connection, because specifying the curvature $G$ and the characteristic class $\h x$ do not specify $C$ completely: one can
still add to $C$ a flat twisted three-form field.  We implicitly complete the definition of an $\mc$ connection in section \ref{formula} by describing
exactly how $C$ couples to an M2-brane, and verify the consistency of the picture in section \ref{consistency}.  But it would be desireable
to have a more succinct definition of an $\mc$ connection.}

Not every manifold admits a $\spinc$ structure or an $\mc$ structure.  In the $\spinc$ case, the obstruction is as follows.
Consider the short exact sequence of groups
\be\label{ares}0\to \Z\overset{2}{\longrightarrow}\Z\overset{r}{\longrightarrow}\Z_2\to 0, \ee
where the first map is multiplication by 2 and the second is reduction mod 2.  This leads to a long exact sequence of cohomology groups
that reads in part
\be\label{wares}\dots \to H^2(X;\Z)\overset{r}{\longrightarrow} H^2(X;\Z_2)\overset{j}{\longrightarrow} H^3(X;\Z)\overset{2}\longrightarrow H^3(X;\Z)\to\dots.\ee
Here $j$ is called the connecting homomorphism, and $j(w_2(X))$ is called $W_3(X)$.  The composition of successive maps in eqn. (\ref{wares}) vanishes, so $2W_3=2\circ j(w_2)=0$ and thus $W_3$ is a 2-torsion class.  The exactness of the sequence (\ref{wares}) says that $w_2$ is the mod
2 reduction of some $\h c_1\in H^2(X,\Z)$ if and only if $W_3=0$.  This is the obstruction to existence of a $\spinc$ structure.
(For a simple example of a manifold in which this obstruction is non-trivial, one may take a $\CP^2$-bundle over $S^1$ such that $\CP^2$ is
complex conjugated in going around $S^1$.)

The obstruction to an $\mc$ structure is similar.  The exact sequence of (\ref{ares}), after being tensored with the orientation bundle of $Y$,
leads to a long exact sequence of twisted cohomology groups of $Y$ that reads in part
\be\label{waress}\dots \to H^4(Y;\ti\Z)\overset{r}{\longrightarrow} H^4(Y;\Z_2)\overset{j}{\longrightarrow} H^5(Y;\ti\Z)\overset{2}\longrightarrow H^5(Y;\ti\Z)\to\dots.\ee
As before, we define $\ti W_5=j(w_4)$; it is a 2-torsion element of $H^5(Y,\ti\Z)$.  Exactness of the sequence tells us that $w_4$ can be lifted to
an integral class $\h x$ if and only if $\ti W_5=0$.  Thus M-theory can be formulated on a $\pin^+$ manifold $Y$ only if\footnote{This
condition has apparently not been stated previously.  However, a somewhat similar condition $W_7=0$ (where as in the previous examples
$W_7=j(w_6)$) was argued in \cite{DMW}.}  $\ti W_5(Y)=0$.  For an example of a $\pin^+$ manifold with $\ti W_5\not=0$, one may take
$\RP^8$ or (to get an eleven-dimensional example) $\RP^8\times \R^3$.  Indeed, $w_4(\RP^8)$ is the nonzero element of $H^4(\RP^8;\Z_2)\cong\Z_2$,
and cannot be lifted to a twisted integral class, since $H^4(\RP^8;\ti \Z)=0$.

\subsection{Formula For The M2-Brane Path Integral Measure}\label{formula}

In this section, we will propose a formula for the path integral measure of an M2-brane wrapped on $W\subset Y$.  
The consistency of this formula will be demonstrated in section \ref{consistency}.

Naively, as in eqn. (\ref{donu}), the path integral measure is $\Pf(\slashed{D})\exp(\i\int_W C)$.  However, both factors have ambiguous signs: $\Pf(\slashed{D})$ has
a sign problem because of the anomaly, and $\exp(\i\int_WC)$ has a sign problem because $C$ is an $\mc$ connection rather than an ordinary three-form
gauge field.  We aim to make sense of the product of these two factors.

To begin with, let us assume that $W$ is the boundary of a submanifold $U\subset Y$ that has the same property as $W$: its normal bundle is oriented.   To
be precise about this, note that once $U$ is given, the normal bundle $N$ to $W$ in $Y$ decomposes as $\o\oplus N'|_W$, where $\o $ is a trivial
real line bundle (generated by the outward normal to $W$ in $U$), $N'$ is the normal bundle to $U$ in $Y$, and $N'|_W$ is the restriction of $N'$ to 
$W$.  Because of this splitting, an orientation of $N$ determines an orientation of $N'|_W$ and the condition we want on $U$ is that this orientation of 
$N'|_W$ extends to an orientation of $N'$.    This condition along with $w_2(Y)=0$ implies the obvious analog of eqns. (\ref{delmo}) and (\ref{belmo}):
\be\label{yelmo} w_1(N')=0,~~~~w_2(N')=w_2(U). \ee
 Given the choice of $U$, we will replace the ill-defined quantities $\exp(\i\int_WC)$ and $\Pf(\slashed{D})$ with
well-defined expressions.  Then
we will argue that the product of these expressions does not depend on the choice of $U$.

Once $U$ is picked, we can simply replace the ill-defined $\exp(\i\int_W C)$ with the well-defined $\exp(\i\int_U G)$.  The motivation for this replacement
is  that the two would be equal if $C$ were an ordinary three-form gauge field, with $G$ obeying standard Dirac quantization.
 As it is, $\exp(\i\int_UG)$ is well-defined for given $U$, and is invariant under small displacements of $U$ because $\d G=0$. But
 $\exp(\i\int_UG)$ does depend on the topological choice of $U$,  because $G$ obeys the shifted quantization condition (\ref{zasess}).

As usual, to make sense of $\Pf(\slashed{D})$ once $U$ is given, we invoke the Dai-Freed theorem \cite{DF}.  This theorem is applicable if we are given
on  $U$ a Dirac
operator $\slashed{D}'$ that is related to $\slashed{D}$ in a suitable fashion along $W=\partial U$.  In the present case, we take $\slashed{D}'$ to be
the natural Dirac operator on $U$ acting on $P|_U$ (that is, on the restriction to $U$ of the $\pin^+$ bundle $P$ of $Y$).   The Dai-Freed theorem
suggests replacing $\Pf(\slashed{D})$ with $|\Pf(\slashed{D})|\exp(-\i\pi\eta/2)$, where here $\eta=\eta(\slashed{D}')$
is the eta-invariant  of the operator $\slashed{D}'$, with APS
boundary conditions along $W=\partial U$.  As usual, the justification for this step is that  $|\Pf(\slashed{D})|\exp(-\i\pi\eta/2)$ is gauge-invariant and well-defined
(once $U$ is picked), depends on the metric only along $W$ and not along $U$, and depends in a  physically sensible\footnote{``Physically sensible''
means that when the metric of $U$ is varied, the variation of $|\Pf(\slashed{D})|\exp(-\i\pi\eta/2)$ depends only on the variation of the metric of $W$.
Moreover, this variation is given 
 in the expected way by a matrix element of the stress tensor. An important consequence is that 
 $|\Pf(\slashed{D})|\exp(-\i\pi\eta/2)$
  varies smoothly with the expected simple zero when the metric variation is such that   $\slashed{D}$ acquires a zero-mode.} way on the metric
of $W$.

The upshot of this is to replace $\Pf(\slashed{D})\exp(\i\int_W C)$ with
\be\label{melf}\ZZ_U=|\Pf(\slashed{D})| \exp(-\i\pi\eta_U/2) \exp\left(\i\int_UG\right). \ee
Here we denote the eta-invariant as $\eta_U$ to emphasize its dependence on $U$.  To show that the formula $\ZZ_U$ for the measure of the M2-brane
path integral makes sense, we have to show that it is independent of the choice of $U$.  There is a standard strategy to prove this, essentially illustrated in fig. \ref{first} of section \ref{condensed}.
We let $U'$ be some other submanifold of $Y$ with boundary $W$, and let $U^*$ be the closed manifold\footnote{For dimensional reasons,
one can assume that $U$ and $U'$ do not intersect away from $W$ so that $U^*$ is embedded in $Y$.  But the following would actually make sense without
this assumption.}
 obtained by gluing $U$ to $-U'$.   Because of the gluing law for the eta-invariant, the desired relation $\ZZ_U=\ZZ_{U'}$ is equivalent to
 \be\label{nelf}\exp(-\i\pi\eta_{U^*}/2)\exp\left(\i\int_{U^*}G\right)=1.\ee
 This can be restated without reference to $G$ using the flux quantization condition (\ref{ozases}):
 \be\label{pelf}\exp(-\i\pi\eta_{U^*}/2)=(-1)^{\int_{U^*}w_4(Y)}.\ee
 This relation will be demonstrated in section \ref{consistency}.
 
 At the outset of this section, we assumed that $W$ is the boundary of some $U\subset Y$.  If this is not so, we are in a situation somewhat like what was considered
 at the end of section \ref{interpretation}.   If $W$ is not a boundary, and thus represents a nontrivial element of the twisted homology $H_3(Y;\ti \Z)$, then the path
 integral measure of an M2-brane wrapped on $W$ cannot be determined from a knowledge of $G$ only and not $C$.  It would be possible to add to $C$ a flat
 three-form gauge field with $\int_WC\not=0$, leaving $G$ unchanged but rotating by a constant phase the path integral measure for an M2-brane wrapped on $W$.
 To understand how to proceed, suppose for example that $H_3(Y;\ti\Z)\cong \Z$ and pick a generator $W_0$.  The phase of the path integral measure for an M2-brane
 wrapped on $W_0$ can be changed in an arbitrary fashion by adding to $C$ a flat three-form gauge field.  Thus we proceed by making an arbitrary choice of this
 phase, and then we attempt to use eqn. (\ref{melf}) to determine the path integral measure for an M2-brane wrapped on any other $W\subset Y$.
 For this, we observe first that, since $W_0$ generates $H_3(Y; \ti\Z)$, there is some integer $n$ such that the disjoint union of $W$ with $n$ copies\footnote{We
 can displace these $n$ copies slightly so that they do not intersect each other or $W$, and for dimensional reasons we can assume that $U$ is embedded in $Y$.
 Once the phase of the path integral measure is fixed for $W_0$, it can be determined for any small displacement of $W_0$ by evolving it continuously, relying
 on the fact that the M2-brane path integral suffers only from global anomalies. An alternative way to determine what happens in a small displacement of $W_0$ 
 is to note that if $W$ can be reached from $W_0$ by a small displacement, this means that the disjoint union of $W$ and $-W_0$ is a boundary in $Y$; the 
 argument in the text will then show that the path integral measure for $W$ is determined in terms of $W_0$.} of $W_0$ is the boundary of some $U\subset Y$.
 This means that the formula (\ref{melf}) determines the path integral measure for a collection of M2-branes consisting of one brane wrapped on $W$ and $n$ others
 wrapped on (slightly separated copies of) $W_0$.  Since the M2-brane path integral measure is known for $W_0$, this determines it for $W$.  
 
 This procedure
 generalizes in a fairly obvious way for any $H_3(Y;\ti \Z)$ and uniquely determines the M2-brane path integral measure for any $W$, up to the possibility of twisting by
 a flat $C$-field.

\subsection{Consistency Of The Formula}\label{consistency}

The statement (\ref{pelf}) that we wish to justify depends only on the four-manifold $U^*\subset Y$ and the normal bundle $N'$ to $U^*$ in $Y$.
Here $Y$ is completely arbitrary except for the constraint that $w_2(Y)=0$, and the only condition on $N'$ is that it is oriented. 
The upshot is that eqn. (\ref{pelf}) must hold with $U^*$ replaced by absolutely any compact four-manifold $X$ (without boundary) and $N'$ replaced
by any oriented rank 7 real vector bundle $B\to X$ that satisfies 
\be\label{donkey}w_1(B)=0, ~~~w_2(B)=w_2(X).\ee
 For $Y$ we can simply take the total space of the vector bundle $B$.
We want to prove eqn. (\ref{pelf}) in this generality. (In what follows, we write just $\eta$ for $\eta_X=\eta_{U^*}$.)

The conditions (\ref{donkey}) ensure that $w_2(TX\oplus B)=0$, and therefore there is a corresponding $\pin^+$ bundle $P$.  
On this $\pin^+$ bundle $P$, there act gamma matrices $\gamma_\mu,~\mu=1,\dots,4$ of $TX$ and additional gamma matrices $\ti\gamma_a$, $a=1,\dots,7$
of $B$, obeying the Clifford algebra
\be\label{delf}\{\g_\mu,\g_\nu\}=2g_{\mu\nu},~\{\ti\g_a,\ti\g_b\}=2\delta_{ab},~\{\g_\mu,\ti\g_a\}=0.\ee  Here $g_{\mu\nu}$ and $\delta_{ab}$ are 
respectively the Riemannian metric of $X$ and the metric of $B$.

Assuming that fermions transform under spatial reflection by the four-dimensional analog of eqn. (\ref{ilb}),
\be\label{nilb}\sR(\psi(x_1,x_2,x_3,x_4))=\pm \g_1\psi(-x_1,x_2,x_3,x_4) ,\ee
the standard Dirac operator $\i\slashed{D}=\i \sum_\mu\g^\mu D_\mu$ (where $D_\mu$ is the connection on $P$) 
is actually odd under a reflection, and the natural self-adjoint Dirac operator
on a possibly unorientable four-manifold is not $\i\slashed {D}$ but
\be\label{wilb}\D=\bg \slashed{D}.\ee
(In going around an orientation-reversing loop in $X$, both $\bg$ and $\slashed{D}$ change sign so $\D$ is well-defined.)  
On an orientable four-manifold, $\D$ is conjugate
to  $\i\slashed{D}$ by $(1+\i\bg)/\sqrt 2$, so the choice between them does not matter, but $\D$ is the version that generalizes to an unorientable four-manifold.
Informally, $\D$ is the Dirac operator on $X$ with values in spinors of $B$, but that is a rough description as in general $w_2(TX)$ and $w_2(B)$ are
nonzero, so that $P$ exists but cannot be decomposed as the tensor product of a spin bundle of $TX$ and one of $B$.

We have written eqn. (\ref{nilb}) in four-dimensional terms, but from an 11-dimensional point of view, $\sR$ may act as a reflection on some of the
normal directions in $B$.  For future reference, we will write the 11-dimensional version of the formula.  If in local coordinates $x_1,x_2,\dots, x_{11}$, a
diffeomorphism
$\rho$ acts as $-1$ on some subset of coordinates $x_{I_1},x_{I_2},\dots,x_{I_s}$, then a section $\Psi$ of $P$ transforms under $\rho$ as
\be\label{filb} \rho\Psi(x)=\pm \G_{I_1}\G_{I_2}\dots \G_{I_s}\Psi(\rho(x)). \ee
Here $\Gamma_I$, $I=1,\dots,11$ are the whole set of gamma matrices $\g_1,\dots,\g_4,\ti\g_1,\dots,\ti\g_7$.  The undetermined sign $\pm$ 
is present as usual because the group that acts on spinors is a double cover of the orthogonal group.

Aiming to prove the identity (\ref{pelf}), our first step is just to give, for any $X$, an example of a bundle $B$ that obeys the conditions (\ref{donkey}).
This example may seem rather special, but it will turn out that studying it will give almost all we need.
To construct such an example, let $TX$ be the tangent bundle of $X$, define
the real line bundle\footnote{Here $\varepsilon$, defined in another way in footnote \ref{uprho}, is a trivial real line bundle on which an orientation-reversing
symmetry acts as $-1$.}
$\varepsilon=\det TX$, and let $\o$ be a trivial real line bundle.  Then an example of an allowed $B$ is
\be\label{dolfo}B= \varepsilon\otimes TX\oplus \varepsilon\oplus \o\oplus \o. \ee
To show that $w_1(B)=0$, $w_2(B)=w_2(TX)$, 
suppose first that  $TX=\oplus_{i=1}^4 \z_i$ is a direct sum of real line bundles $\z_i$; this implies that $\varepsilon
=\otimes_{i=1}^4 \z_i$.   In this case, one can verify the desired relations $w_1(B)=0$, $w_2(B)=w_2(TX)$
 using the Whitney sum formula (\ref{sumf}) and the relation $w_1(\otimes_{\alpha} \z_\alpha)=\sum_\alpha w_1(\z_\alpha)$, valid for an arbitrary
collection of real line bundles.  The ``splitting formula'' for characteristic classes of vector bundles implies that the general result we want follows
from this seemingly very special case.

For this particular $B$, the $\pin^+$ bundle of $TX\oplus B$ can be described in a particularly simple and useful way.  But this will take a number of steps.
Let us begin with the case that $X$ is orientable, so (ignoring for now the action of orientation-reversing diffeomorphisms)
we can set $\varepsilon=\o$ and $B$ becomes $B'\oplus \o^{\oplus 3}$, with $B'=TX$.
Moreover, to start with, let us omit the trivial summand $\o^{\oplus 3}$ and just consider the spinors of $TX\oplus B'=TX\oplus TX$.
Thus, the Clifford algebra of $TX\oplus B'$ in this special case is just the product of two copies of the Clifford algebra of $TX$.
So we have two sets of  gamma matrices $\g_\mu$ and $\ti\g_\mu$, $\mu=1,\dots,4$, and the algebra (\ref{delf}) becomes
\begin{align}\label{udelf}\{\g_\mu,\g_\nu\}&=\{\ti\g_\mu,\ti\g_\nu\}=2g_{\mu\nu}, \cr \{\g_\mu,\ti\g_\nu\}&=0.\end{align}

It is fairly well known that the doubled Clifford algebra (\ref{udelf}) is related to differential forms on $X$, but there is a small twist that may come
as a surprise.  On differential forms on $X$, there acts a Clifford algebra that is generated, in a local coordinate system, by $\d x^\mu$ (that
is by the wedge product with the one-form $\d x^\mu$) and the corresponding contraction operator $\iota_{\d x^\mu}$.  They obey
\be\label{weelf}\{\d x^\mu,\d x^{\nu}\}=0=\{\iota_{\d x^\mu},\iota_{\d x^\nu}\},~~\{\d x^\mu,\iota_{\d x^\nu}\}=\delta^\mu_\nu.\ee
and this algebra acts irreducibly on the differential forms $\Omega^*(X)=\oplus_{j=0}^4 \Omega^j(X)$ (here $\Omega^j(X)$ is the bundle of $j$-forms
on $X$).
We can reproduce this algebra from the doubled Clifford algebra (\ref{udelf}) by setting
\be\label{eelf}\d x^\mu=\frac{1}{\sqrt 2}\left(\i\g^\mu+\ti\g^\mu\right),~~~\iota_{\d x^\nu}=\frac{1}{\sqrt 2}\left(-\i\g_\nu+\ti\g_\nu\right). \ee
Thus, the doubled Clifford algebra (\ref{udelf}) acts on the space $\Omega^*(X)$ of differential forms on $X$.  

However, as the doubled Clifford
algebra (\ref{udelf}) and the corresponding algebra generated by $\d x^\mu$ and $\iota_{\d x^\nu}$ are both real, the factors of $\i=\sqrt{-1}$
that appear in eqn. (\ref{eelf}) are unnatural.  Technically, those factors mean that by this mapping, we can identify the complexification
of $\Omega^*(X)$ with a complexification of a  spin bundle of $TX\oplus TX$.   

We can do better if we introduce an orientation of $X$ and introduce the chirality operators $\bg=\g_1\g_2\g_3\g_4$, $\bg'=\ti\g_1\ti\g_2\ti\g_3\ti\g_4$
of the first and second summands of $TX\oplus TX$.
  We replace $\i$ with $\bar\g$ so that (\ref{eelf}) becomes
  \be\label{yelf}\d x^\mu=\frac{1}{\sqrt 2}\left(\bg\g^\mu+\ti\g^\mu\right),~~~\iota_{\d x^\nu}=\frac{1}{\sqrt 2}\left(-\bg\g_\nu+\ti\g_\nu\right). \ee
This map is real and is simply an isomorphism between the doubled Clifford algebra of $TX$ and the natural Clifford algebra that acts on $\Omega^*X$.

Up to this point, there is a different isomorphism that we could have made instead:
\be\label{yelff}\d x^\mu=\frac{1}{\sqrt 2}\left(\g^\mu+\bg'\ti\g^\mu\right),~~~\iota_{\d x^\nu}=\frac{1}{\sqrt 2}\left(\g_\nu-\bg'\ti\g_\nu\right). \ee
This also gives an isomorphism of the Clifford algebra of $TX\oplus TX$ with the algebra generated by $\d x^\mu$ and $\iota_{\d x^\nu}$.  
However, the mapping (\ref{yelf}) is the one we want, because under this mapping, the Dirac operator $\D$ defined in eqn. (\ref{wilb}) maps
to the natural Dirac-like operator $\d+\d^*=(\d x^\mu-g^{\mu\nu}\iota_{\d x^\nu})D_\mu$ acting on $\Omega^*(X)$.

An important operator on $\Omega^*(X)$ is the operator $(-1)^\sF$ that takes the value $(-1)^d$ on a differential form of degree $d$.  This operator
anticommutes with $\d x^\mu$ and $\iota_{\d x^\mu}$, and thus with all the gamma matrices.  So it is the overall chirality operator
\be\label{dolly}\bG=\bg\,\bg'=\g_1\g_2\g_2\g_4\g_1'\g_2'\g_3'\g_4'.\ee
(In a moment, it will be important that this operator can be defined globally even if $X$ is unorientable.  That is because $w_1(TX\oplus \varepsilon\otimes TX)=0$.)

As a step toward considering what happens if $X$ is unorientable, we will now analyze the effects of an orientation-reversing diffeomorphism of $X$
in the above story.   With this in view, we return to eqn. (\ref{dolfo}) and note that the first summand $B'$ of $B$ is really $\varepsilon\otimes TX$ rather than $TX$.
Here $\varepsilon$ is a trivial line bundle if $X$ is orientable, so we have omitted it so far, but an orientation-reversing diffeomorphism of $X$ acts
as $-1$ on $\varepsilon$, so it will be important henceforth to include this factor.  Also, 
 it will be important to include the additional summand $\varepsilon$ of
$B$.  Thus when $X$ is possibly unorientable, we will compare $\Omega^*(X)$ to a $\pin^+$ bundle of $TX\oplus B''$, where $B''=B'\oplus\varepsilon=\varepsilon\otimes TX
\oplus \varepsilon$.

For the moment, however, we still assume that $X$ is orientable and describe the spin bundle of $TX\oplus B''=TX\oplus \varepsilon\otimes TX\oplus \varepsilon$.
To do this, we need to extend the Clifford algebra that we used before by adding a ninth generator $\g_\varepsilon$, corresponding to the summand $\varepsilon$ of $B''$.
This ninth gamma matrix must anticommute with the eight that we already have, so it must be\footnote{We could reverse the sign of $\g_\varepsilon$,
but anyway we will be including both signs momentarily.} $\g_\varepsilon=\bG$, where $\bG$ was defined in eqn. (\ref{dolly}).  Note
that we can add this new generator to the Clifford algebra without changing the bundle that the algebra acts on.

Next we  compare the action of an orientation-reversing diffeomorphism on a spin bundle of $TX\oplus B''$ to its action on $\Omega^*(X)$.
Let us consider an orientation-reversing reflection of $X$ that in local coordinates acts as
\be\label{mello}\rho(x_1,x_2,x_3,x_4)= (-x_1,x_2,x_3,x_4), \ee
and acts accordingly on $B$.
By virtue of eqn. (\ref{filb}), the element of the Clifford algebra that will represent this reflection is
$\uprho=\g_1\ti\g_2\ti\g_3\ti\g_4\g_\varepsilon$.   (Here we include $\g_1$ and not $\g_i$ for $i=2,3,4$ because only $x_1$ is being reflected;
for the $\ti\g_\mu$, we make the opposite choice and include $\ti\g_i$, $i=2,3,4$ because these are gamma matrices of $\varepsilon\otimes TX$
where the reflection acts as $-1$ on $\varepsilon$; and likewise because $\varepsilon$ is odd under the reflection, we include $\g_\epsilon$.)   With $\g_\varepsilon=\bG$,
we get $\uprho=\g_2\g_3\g_4\ti\g_1$.  Using the definitions (\ref{yelff}), this means that $\uprho$ anticommutes with $\d x^1$ and $\iota_{\d x^1}$, and commutes
with $\d x^i$ and $\iota_{\d x^i}$ for $i>1$.  This is exactly the expected action of $\uprho$ on differential forms.

Thus, the identification betwen the  spin bundle of $TX\oplus B''$ and $\Omega^*(X)$ maps the action of an orientation-reversing diffeomorphism
on one side to the action of such a diffeomorphism on the other side.    Now we are finally ready to state a similar identification for the case that $X$
is not necessarily orientable.

Suppose that  the unorientable manifold $X$ is the quotient of an orientable
manifold $X'$ by a $\Z_2$ symmetry generated by a freely acting orientation-reversing diffeomorphism 
$\rho$.  After identifying as before  the spin bundle of $TX'\oplus \varepsilon\otimes TX'\oplus \varepsilon$
 with $\Omega^*(X')$, and then dividing by $\rho$ on both sides,
we learn that as promised we can identify a $\pin^+$ bundle of $TX\oplus B''$ with $\Omega^*(X)$.

To be more precise, this argument shows that there is a particular $\pin^+$ bundle of $TX\oplus B''$ that can be identified with $\Omega^*(X)$.
Any other $\pin^+$ bundle is obtained by twisting this one with a real line bundle $\upalpha$ over $X$.  Thus a general $\pin^+$ bundle of 
$TX\oplus B''$ is what we might call $\Omega_\upalpha^*(X)$, the differential forms on $X$ twisted by the real line bundle $\upalpha$.  For example,
if as before we construct $X$ as $X'/\Z_2$ where $\Z_2$ is generated by $\rho$,
then starting with the differential forms $\Omega^*(X')$ and dividing by $\Z_2$, the two choices of sign in the action of $\rho$ on the $\pin^+$ bundle $\Omega^*(X')$ will
give us on the quotient either $\Omega^*(X)$ or $\Omega_\varepsilon^*(X)$, the differential forms on $X$ twisted by $\varepsilon$. 

Finally, let us discuss the effect of including the last two summands $\o\oplus\o$ in (\ref{dolfo}).  To do this, we have to double the rank of the $\pin^+$ bundle.
There is a straightforward construction.  Let $\Gamma'_I$, $I=1,\dots,9$ be the whole set of gamma matrices of $TX\oplus B''$ (thus the $\Gamma'_I$ are
$\g_\mu$ and $\ti\g_\mu$ for $\mu=1,\dots,4$ and $\g_\varepsilon$).   For the full set of 11 gamma matrices of $TX\oplus B$, we can take
\be\label{zonx}\G_I=\G_I'\otimes \begin{pmatrix}1&0\cr 0&-1\end{pmatrix},  ~I=1,\dots,9,~~ ~~\G_{10}={\mathbf I} \otimes \begin{pmatrix}0&1\cr 1&0\end{pmatrix},~~
\G_{11}={\mathbf I} \otimes \begin{pmatrix}0&-\i\cr \i&0\end{pmatrix}.\ee
(Here $\mathbf I$ is the identity matrix.  It is not possible to choose all 11 gamma matrices to be real, as the spinors of $\Spin(11)$ are pseudoreal.)
If we ignore the last two gamma matrices, which do not enter the construction of the Dirac operator, this tells us that the $\pin^+$ bundle of
$TX\oplus B$ is the direct sum of two $\pin^+$ bundles of $TX\oplus B''$ except that in one copy the gamma matrices of $TX\oplus B'$ all  have opposite sign.  What is the effect of this
sign reversal?
Reversing the signs of $\g_\mu$ and $\ti\g_\nu$, $\mu,\nu=1,\dots,4$, is not very consequential since one can compensate for this by conjugating by $\g_\varepsilon$.
But changing the sign of $\g_\varepsilon$ reverses the sign of the element of the Clifford algebra that represents an orientation-reversing symmetry.
(For instance, in the example considered earlier, reversing the sign of $\g_\varepsilon$ changes the sign of $\uprho=\g_1\ti\g_2\ti\g_3\ti\g_4\g_\varepsilon$.)
This has the effect of replacing $\Omega^*(X)$ by $\Omega^*_\varepsilon(X)$ and more generally for any real line bundle $\upalpha$
of replacing $\Omega^*_\upalpha(X)$ by
$\Omega^*_{\upalpha\otimes\varepsilon}(X)$.

The upshot then is that one particular $\pin^+$ bundle $P$ of $TX\oplus B$ is equivalent to $\Omega^*(X)\oplus \Omega^*_\varepsilon(X)$
and a general $\pin^+$ bundle $P_\upalpha$ of $TX\oplus B$, obtained by twisting $P$ by a real line bundle $\upalpha$,
 is $\Omega^*_\upalpha(X)\oplus \Omega^*_{\upalpha\otimes\varepsilon}(X)$.  
 
 Now we can calculate the eta-invariant
and verify the identity (\ref{pelf}).
We use the fact  that the Dirac operator $\D$ acting on $P$ reduces to the operator $\d+\d^*$ acting on $\Omega^*(X)\oplus \Omega^*_\varepsilon(X)$
(or a version of this twisted by some $\upalpha$, as we discuss shortly).  Moreover, $\d+\d^*$ anticommutes with the operator $(-1)^\sF$ that measures the degree of
a differential form mod 2.  Accordingly, the spectrum of the operator $\d+\d^*$ is invariant under $\lambda\leftrightarrow-\lambda$.  This implies that nonzero
eigenvalues of $\D$ make no net contribution to the eta-invariant.  This invariant is therefore simply the number of zero-modes of $\D$.

Let $b_i$, $i=1,\dots,4$  be the $i^{th}$ Betti number of $X$ (the number of $i$-form zero-modes of $\d\oplus \d^*$ acting on $\Omega^*(X)$).  The total
number of zero-modes of $\d+\d^*$  is $\sum_{i=1}^4b_i$, and this is the eta-invariant of $\D$ acting on $\Omega^*(X)$.  When $\d\oplus \d^*$ acts on 
$\Omega_\varepsilon^*(X)$,  the number of $i$-form zero-modes is\footnote{\label{hodge} On an orientable manifold,  the Hodge $\star$ operator  establishes
an isomorphism between $\Omega^i(X)$ and $\Omega^{4-i}(X)$, leading to an identity $b_i=b_{4-i}$.  However, because $\star$ is odd under reversal
of orientation, on an unorientable manifold, it maps $\Omega^i(X)$ to $\Omega_\varepsilon^{4-i}(X)$, showing that the number of $i$-form zero-modes
of $\d+\d^*$ acting on $\Omega^*(X)$ is the number of $(4-i)$-form zero modes of $\d+\d^*$ acting on $\Omega_\varepsilon^*(X)$, and vice-versa.}
$b_{4-i}$, and the total number of zero-modes is again $\sum_{i=1}^4b_i$.
Thus the eta-invariant of $\D$ acting on the $\pin^+$ bundle $\Omega^*(X)\oplus \Omega^*_\varepsilon(X)$ is $\eta=2\sum_{i=1}^4 b_i$.

Let us compare this to the Euler characteristic of $X$, defined as $\chi=\sum_{i=0}^4 (-1)^i b_i$.  We see immediately that 
\be\label{zetatt}\eta=2\chi~~{\mathrm{mod}}~4.\ee
Therefore we can evaluate the left hand side of 
the identity (\ref{pelf}) that we are trying to verify:
\be\label{etat}\exp(-\pi\i\eta/2)=(-1)^\chi.\ee
To evaluate the right hand side of this identity, we let $\hat\chi$ be the Euler class of $TX$.  The Euler characteristic of $X$ is $\chi=\int_X\hat\chi$,
and on the other hand the mod 2 reduction of $\hat\chi$ is the Stieffel-Whitney class $w_4$.  So $(-1)^\chi=(-1)^{\int_Xw_4}$ and we get the desired
identity,
\be\label{zetat}\exp(-\pi\i\eta/2)=(-1)^{\int_X w_4}.\ee

It is straightforward to extend this computation to the general case that $\D=\d+\d^*$ acts on $\Omega_\upalpha^*(X)\oplus \Omega^*_{\upalpha\otimes \varepsilon}$ for some
$\upalpha$.
Let $b_{i,\upalpha}$ be the number of zero modes of $\d+\d^*$ acting on $\Omega_\upalpha^*(X)$.  One might call these the Betti numbers of the $\upalpha$-twisted
cohomology.  Then $b_{i,\upalpha\otimes \varepsilon}=b_{4-i,\upalpha}$ as in footnote \ref{hodge}.  Just as before, $\eta=2\sum_i b_{i,\upalpha}$
and hence is congruent mod 4 to the Euler characteristic of the $\upalpha$-twisted cohomology, which is defined as  $\chi_\upalpha=\sum_{i=0}^4(-1)^ib_{i,\upalpha}$.  However, the Gauss-Bonnet
formula expressing the Euler characteristic as a curvature integral is valid for any $\upalpha$ and shows that $\chi_\upalpha$ is independent of $\upalpha$.
Hence $\exp(-\pi\i\eta/2)$ is independent of $\upalpha $ and the identity (\ref{zetat}) holds for all $\upalpha$.

So far we have shown that for any $X$, there is some choice of  $B$ -- described in eqn. (\ref{dolfo}) -- such that the identity
(\ref{zetat}) holds.  To complete the argument, we will now show the following:  for a given $X$, the identity (\ref{zetat}) holds for one choice
of $B$ if and only if it holds for any $B$.   In proving this, we will use the fact that the left and right hand sides of eqn. (\ref{zetat}) are both cobordism
invariant.  Cobordism invariance means that they equal 1 if $X$ is the boundary of a five-manifold $Z$ that is endowed with appropriate structure.
Concerning the right hand side of eqn. (\ref{zetat}), no additional structure is needed;
 the cobordism invariance of $(-1)^{\int_X w_4}$ is a special case of cobordism invariance of Stieffel-Whitney
numbers (integrals of products of Stieffel-Whitney classes).  On the left hand side, we are given $X$ and $B$ and a $\pin^+$ structure of $TX\oplus B$.
We have to require that the rank 7 real vector bundle $B$ over $X$ extends to a rank
7 real vector bundle $B_Z$ over $Z$, such that there is a $\pin^+$ structure of $TZ\oplus B_Z$ that restricts on the boundary to the given $\pin^+$ structure of
$TX\oplus B$.   (These conditions are straightforwardly satisfied in the example
considered in the next paragraph.)    Under these circumstances, we can define a Dirac operator $\D_Z$ on $Z$ that is related to the Dirac operator $\D$ of $X$ in
a way that will let us invoke the Atiyah-Patodi-Singer index theorem \cite{APS}. Let $\eta$ be the eta-invariant of the Dirac operator
on $X$ and $\I$ the index\footnote{On a $\pin^+$ manifold of odd dimension, the Dirac operator maps sections of one $\pin^+$ bundle to sections
of the complementary $\pin^+$ bundle.  Thus it is not self-adjoint (that is why the anomaly studied in the three-dimensional case in section \ref{analysis} is
possible) and in particular can have a nonzero index.   This index vanishes on a compact manifold without boundary (this statement is a special case of the ordinary
Atiyah-Singer index theorem), but on a manifold with boundary, with APS boundary conditions, the index can be nonzero.  For example, see Appendices B.4 and C
of \cite{Witten}.} of $\D_Z$.  The APS index theorem
 says that $\eta/2=\I$, but on the other hand $\I$ is even because of a version of Kramers doubling.   So $\eta$ is a multiple of 4 and $\exp(-\pi\i\eta/2)=1$,
establishing the cobordism invariance of the left hand side of eqn. (\ref{zetat}).

\begin{figure}
 \begin{center}
   \includegraphics[width=3.5in]{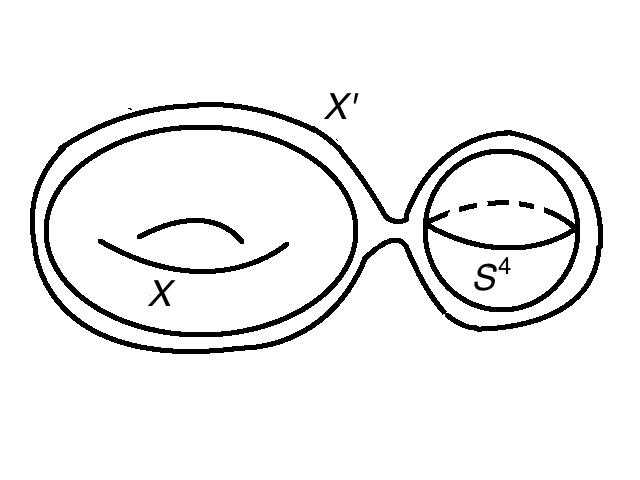}
 \end{center}
\caption{\small  Like peas in a pod, the disjoint union of a four-manifold $X$ and a four-sphere $S^4$ is cobordant to their connected sum $X'$.
The connected sum is defined by cutting an open ball out of each and gluing them together along their boundaries. The normal bundles are trivialized in the cutting
and gluing region before this is done.}
 \label{cob}
 \end{figure}

Now we will use cobordism invariance to show that for a given $X$,  the identity (\ref{zetat}) is true for all $B$ if it is true for one $B$.
Given a triangulation of $B$, the constraints $w_1(B)=0$, $w_2(B)=w_2(X)$ determine $B$ on the two-skeleton of $X$.  Since 
$\pi_2(\SO(7))=0$, there is no further choice to be made on the three-skeleton of $X$.  So any possible $B$ differs from the particular choice in eqn. (\ref{dolfo}) at most by a local modification made near a point $p\in X$.  The local modification
is made by gluing in a certain number of instantons.\footnote{If $X$ is unorientable -- which is the most relevant case as the identity of interest was already
proved in \cite{Wittenold} for orientable $X$ -- then the number of instantons that are glued in is only an invariant mod 2.  (An instanton that is transported around
an orientation-reversing loop in $X$ comes back as an anti-instanton.) So for unorientable $X$, there are actually only two possible choices of $B$ up to 
isomorphism -- the one in (\ref{dolfo}), and one with a different value of $w_4$, reached by gluing in a single instanton. The proof given in the text does not require a knowledge of this fact.}   These are ordinary gauge theory instantons, classified by $\pi_3(\SO(7))=\Z$, where $\SO(7)$ is the structure
group of $B$.   To use these facts, start with a copy of $X$, endowed with some normal bundle $B$ for which
we know that the identity (\ref{zetat}) holds, and a copy
of $S^4$, endowed with a normal bundle of some instanton number $n$.  Since $S^4$ is orientable, the identity (\ref{zetat}) holds for $S^4$ with any normal
bundle by virtue of the computation
in \cite{Wittenold} (the basic idea was explained in section \ref{zerob}).  
Now we form the connected sum of $X$ and $S^4$, by cutting a small open
ball out of each and gluing them together along the resulting boundaries.\footnote{The normal bundle $B$ to $X$ in $Y$ and the analogous normal bundle to $S^4$
are trivialized along these small balls before the cutting and gluing is done.}  The connected sum is a manifold $X'$ that is topologically equivalent to
$X$, but with a local modification of $B$ that shifts the instanton number by $n$.  The identity we want is true for $X$ and for $S^4$ and therefore for their disjoint union,
and $X'$ is cobordant to the disjoint union of $X$ and $S^4$ (fig. \ref{cob}).   So the identity is true for $X'$.  Notice that this argument works for any choice of the $\pin^+$ structure
of $TX\oplus B$.

It is possible to make a similar argument using the gluing theorem for the eta-invariant, rather than its cobordism invariance.

\section{Summary}\label{conclusions}

In section \ref{analysis} of this paper, we extended the standard analysis of the ``parity'' anomaly for fermions
in a spacetime of odd dimension
 to the case of a possibly unorientable manifold.  We gave a general description of the anomaly, valid in any
 dimension, 
in terms of the $\eta$-invariant in one dimension more.  In the important case of $2+1$ dimensions, we made
this formula more concrete for gauge theories with gauge group $\Z_2$, $\U(1)$, or $\SU(2)$, and described
methods by which this could be done for any gauge group.

A key result of section \ref{analysis} was to show that the usual ``parity'' anomaly
on an orientable manifold is not the whole story, and that a theory which is anomaly-free if considered on
an orientable manifold only may be anomalous if formulated on an unorientable manifold.  

The rest of the paper was devoted to two applications.  Section \ref{gb} was devoted to an application to topological
superconductors.  Gapped boundary states of topological insulators and superconductors can be constructed
by postulating suitable emergent gauge fields, matter fields, and symmetry breaking patterns on the surface
of a material.   This approach was developed in \cite{SW}, mainly but not entirely in the context of a topological
insulator.  To fully exploit this approach in the case of a topological superconductor requires a more
complete understanding of the ``parity'' anomaly than was available previously.  In section \ref{gb}, we use
the results of section \ref{analysis} to analyze more fully and precisely the gapped boundary states of a topological superconductor that can be constructed
using these methods.   A typical application is to show that the boundary of a topological 
superconductor with $\nu=16$, but not
one with $\nu=8$, can be gapped in a topologically trivial way (that is, without introducing topological order 
on the boundary).

The existing literature on gapped boundary states of topological insulators and superconductors is
extensive and varied; see for example \cite{WS,MCFV,FCV, KitTwo, MKF,TPfaffian,TPfaffiantwo,Wangetal,Wangetals,Al,VS,
MetlitskiTSC}.  One difference between the present paper (and the previous one \cite{SW}) and some other approaches is that our treatment is completely explicit, based on weak coupling, with no reliance on non-explicit dynamics.\footnote{There is one partial exception to this.  In sections \ref{sixteen} and \ref{confinement}, we describe three
ways to gap the boundary of a $\nu=16$ topological superconductor in a without introducing topological order.  One of the
three approaches relies on standard but non-explicit dynamics, namely confinement in $\SU(2)$ gauge theory without
matter fields in $2+1$ dimensions.}   

The explicit nature of the dynamics has made it possible to see in \cite{SW} and in the present paper an important
feature of these boundary states.  This is a group extension: the bulk relation $\sT^2=(-1)^F$ is generalized
along the boundary to $\slT^2=(-1)^F\sK$, where $\sK$ generates a discrete group of unbroken gauge symmetries.
It is not clear to the author to what extent this has been understood in the previous literature.
In section \ref{bosonic}, we show that an analogous group extension gives a unified way to construct gapped
boundary states of bosonic SPT states.  The presentation in this section is somewhat abstract, but this will hopefully
be remedied in forthcoming work.

Section \ref{mtheory} applies similar methods to a problem in M-theory.  M-theory describes among other things
membranes, usually called M2-branes, with $2+1$-dimensional worldvolume.  It has been known for some time
\cite{Wittenold} that conxistency of the M2-brane path integral depends on a fairly subtle cancellation of the
``parity'' anomaly.  However, this cancellation has hitherto been understood only for the case that the M2-brane
worldvolume is orientable.  With the help of the tools developed in section \ref{analysis}, the unorientable case
is treated in section \ref{mtheory}.

\def\U{{\mathcal U}}
Research supported in part by NSF Grant PHY-1314311.   I thank D. Freed, G. W. Moore, J. Morgan, N. Seiberg, and J. Wang for discussions.
\bibliographystyle{unsrt}

\end{document}